\documentclass[review]{elsarticle}
\usepackage{lineno,hyperref}
\usepackage{subfigure}
\usepackage{color}
\modulolinenumbers[1]
\usepackage{graphicx}
\journal{Bull. Pol. Acad. Sci., Tech. Sci.}
\begin{document}
\title{A new generator of chaotic bit sequences with mixed-mode inputs}
\author{M. Melosik}
\address{Poznan University of Technology, Department of Computer Science\\ Piotrowo 3A, 61-138  Poznan, POLAND}
\ead{michal.melosik@put.poznan.pl}

\author{W. Marszalek\fnref{myfootnote}}
\address{Rutgers University,
Department of Mathematics\\ 110 Frelinghuysen Rd.,
Piscataway, NJ 08854, USA}
\ead{w.marszalek@rutgers.edu}
\fntext[myfootnote]{Corresponding author}

%
%
%

\begin{abstract}
\looseness=-1 This paper presents a   new generator of chaotic bit sequences with mixed-mode (continuous and discrete)  inputs. The generator has an improved level of chaotic properties in comparison with the existing single source (input) digital chaotic bit generators. The 0--1 test is used to show the improved chaotic behavior of our generator having a chaotic continuous input (Chua, R\"{o}ssler or Lorenz system) intermingled with a discrete input (logistic, Tinkerbell or Henon map) with various parameters. 
The obtained sequences of chaotic bits show some features of random processes with
increased entropy levels, even in the cases of small numbers of bit representations. 
 The properties of the new generator and its binary sequences compare well with those obtained from a truly random binary  reference quantum generator, as evidenced by the results of the \textit{ent} tests. 
\end{abstract}

\maketitle

\section{Introduction}\label{sec1}

Most well-known chaotic-based  bit generators  use a single source (input), either a continuous or discrete one \cite{1,2}.
 Such an approach may result in  a bit sequence  that can be compromised through a synchronization of  chaotic systems  \citep{3}-\citep{5},
 or a possible prediction of  a bit sequence due to a finite precision number representation \citep{6}-\citep{8}.
  Those problems are strongly linked to the fact that the output of a single discrete input generator in the finite precision arithmetics becomes periodic (therefore nonchaotic) even if the length of the output sequence is of order $10^6$. Most studies on digital chaotic generators utilize the NIST\footnote{NIST = National Institute of Standards and Technology} tests when the statistical properties (i.e. randomness of the sequences of $0s$ and $1s$) are examined for the sequences of minimum length of $10^6$. Our paper differs from such an approach, as we are interested in improving the level of chaotic behavior (as measured by the parameter $K$ below).

\subsection{Historical perspective}
The best approach to generate random binary sequences is to  use a source with the highest possible entropy level.  That level depends primary on the source itself, its ``working" environment and possible unauthorized attempts to modify the source, for example through hacking. Fig. 1a shows a general idea of generating a random binary sequence. A chaotic bit generator serves as a source of a high entropy sequence. The binary sequences obtained from chaotic generators are called the \emph{binary chaotic sequences}. To transform such sequences into binary random sequences it is necessary to post-process the chaotic sequences and apply corrections to eliminate instances of biased $0$ and $1$ bits (long intervals of bits of the same value). Typical method applied for that purpose is the von Neumann correction that is considered to be a randomness extractor \citep{9}. 

In the correction process, the random bits of a
sequence form pairs  of two bits.
Whenever the two bits of a pair are equal, the pair  
is discarded. When the two bits are different and the
pair starts with a 1, the pair is
replaced by a 1. When it starts with a 0, it is replaced
by a 0. After such a correction, the bias is removed from
the sequence.
  The von Neumann correction transforms a chaotic sequence into a random one, which is shorter than the orignal chaotic sequence. Most of the methods of examining binary sequences are applied to the post-processed sequences, after the von Neumann correction had been applied. This paper takes a different approach and the binary chaotic sequences are analyzed and tested without being pre-processed. This allows for a direct analysis of the obtained bit sequences. The new mixed-mode input  generator of chaotic bit sequences is tested with the 0--1 test for chaos to monitor the generator's chaotic nature. Short bit sequences, apart from their chaotic dynamics, also have selected properties of random processes, confirmed through a compution  of typical parameters, such as,   
  the entropy level, compression, $\chi ^2$ and Monte-Carlo $\pi$ numbers, and others. The sequences obtained from our mixed-mode input chaotic binary generator are compared with a reference binary sequence with high entropy level obtained from the Quantis generator \citep{10}.

\subsection{Other issues of binary sequence generators}
 Increasing the number of bits used in the finite precision arithmetics may  temporarily solve the problem of having a lower quality random sequence. However, the problem repeats again with the increased length of the output sequence, beyond the $10^6$ order. In practical terms, increasing the length of the output sequence, say from order $10^6$ to $10^7$ does not solve the problem, since the longer sequence will eventually also  become periodic \citep{6}.      In order to provide a solution to such problems we designed a new generator of chaotic bits with mixed-mode inputs. Such an approach increases  the level of chaos, as it is much more difficult to predict the chaotic binary output  of our generator. The parameters of the two (continuous and discrete) chaotic input systems can easily be modified each time the generator is  run. It is shown that in the case of small number of bit representation the proposed generator works much better when  compared with the single source chaotic bit generators. Our generator prevents periodicity and the obtained sequence is truly chaotic as evidenced by the results of the 0--1 test. The tests for randomness also show an improved quality (i.e. higher entropy levels) of the binary sequences obtained from the mixed-mode generator.

Also, the proposed generator with two parts, the  discrete (logistic, Henon or other map) and analog (Chua,  R\"{o}ssler or Lorenz system), can be modeled and analyzed using the hardware description language VHDL-AMS, which is an industry standard modeling language for mixed signal circuits. In this paper, however, we focus on analyzing dynamical properties and features of the proposed generator rather than its hardware implementations. The hardware realizations of the proposed generator will be the focus of further research.

\section{New mixed-mode generator of chaotic bit sequences}
\label{sec:1}
Fig. 1b shows the structure of the new generator. The continuous input comes from a chaotic circuit of Chua, R\"{o}ssler, Lorenz (or other), while the discrete input can be a logistic, Henon, Tinkerbell or Baker map. An example of a digital implementation of the logistic map in a finite precision number representation is shown in Fig. 1c. The logistic equation is: $x(n+1)=\mu x(n)[1-x(n)]$ with $1<\mu <4$ and $0<x(0)<1$.

 The continuous input in Fig. 1b is discretized and synchronized with the internal clock of the discrete chaotic map. The two binary signals $\{D_i\}$ and $\{C_j\}$ are mixed through the XOR operation. As a result, a sequence of bits $\{N_k\}$ is obtained. The 0--1 test is then applied to the sequence $\{N_k\}$ to check its chaotic nature. If the result of the test is satisfactory (see more details below), the $\{N_k\}$ sequence is used to form a new sequence of chaotic bits. Otherwise, a change of the parameters of the discrete chaotic map should be considered to yield a better test result. Another option to consider is to replace the discrete chaotic system by a different one. For example, one may consider replacing logistic map by Tinkerbell, Henon, Baker or other discrete map.

\begin{figure*}[!t]
 \centering
\includegraphics[width=0.9\textwidth,height=1.02in]
 {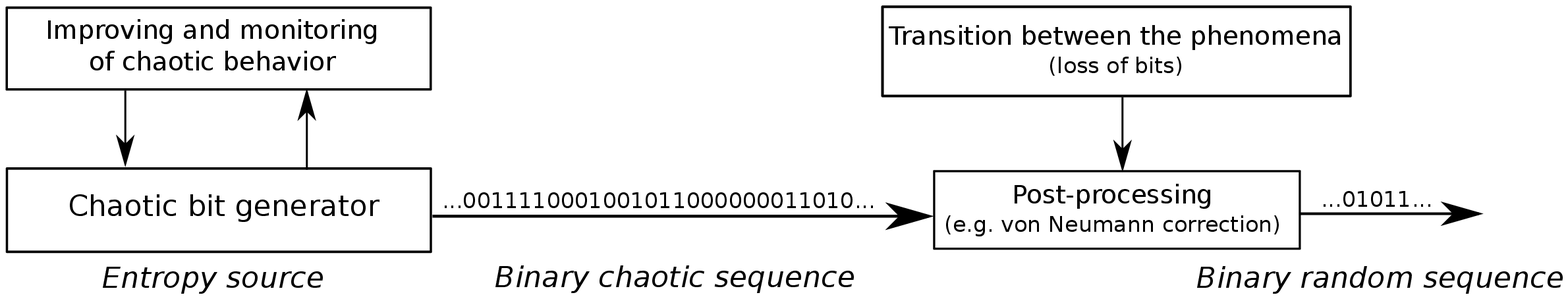}\\ (a)\\
 \includegraphics[width=2.35in,height=2.6in]
 {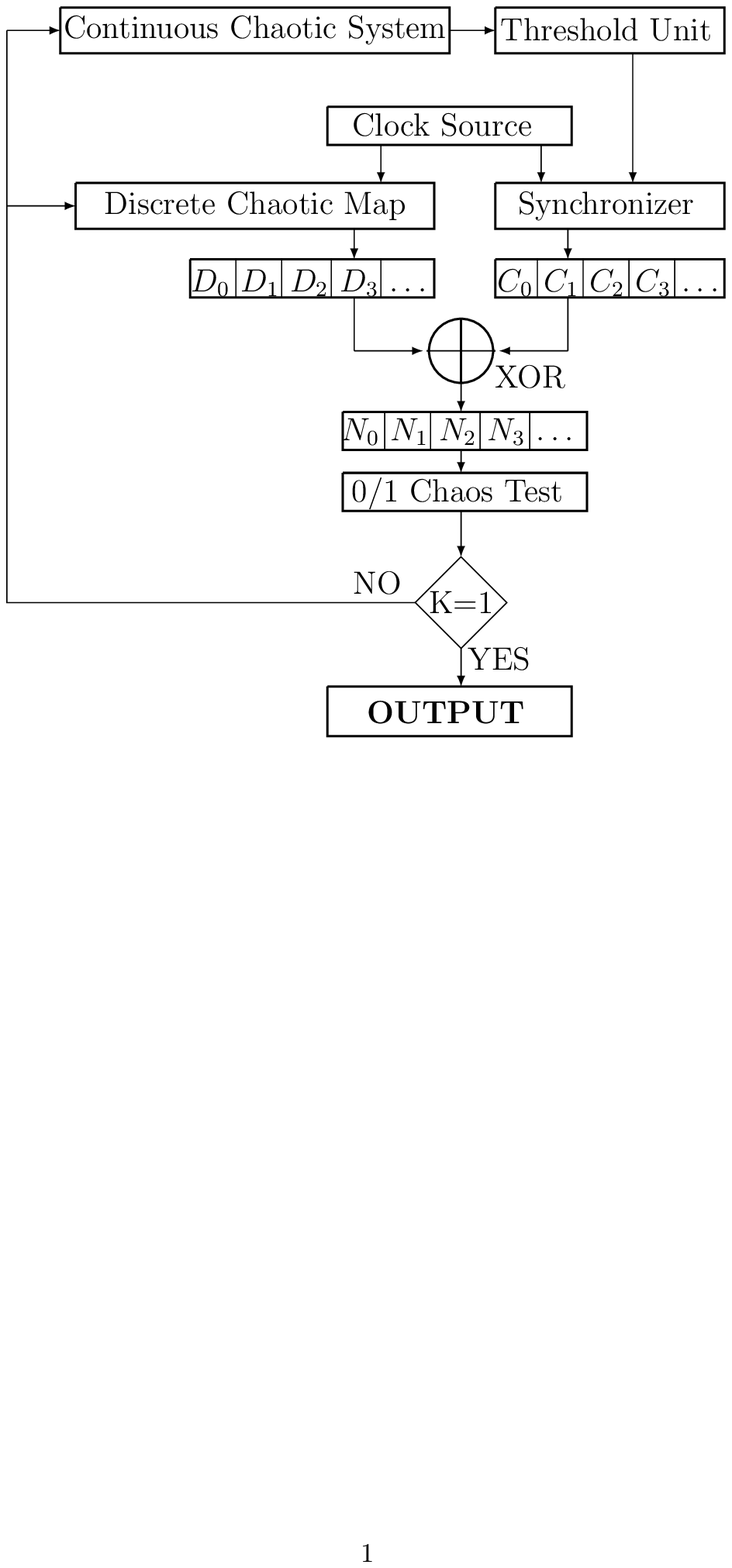}\hspace{0.5cm}
 \includegraphics[width=2.1in,height=2.15in]
 {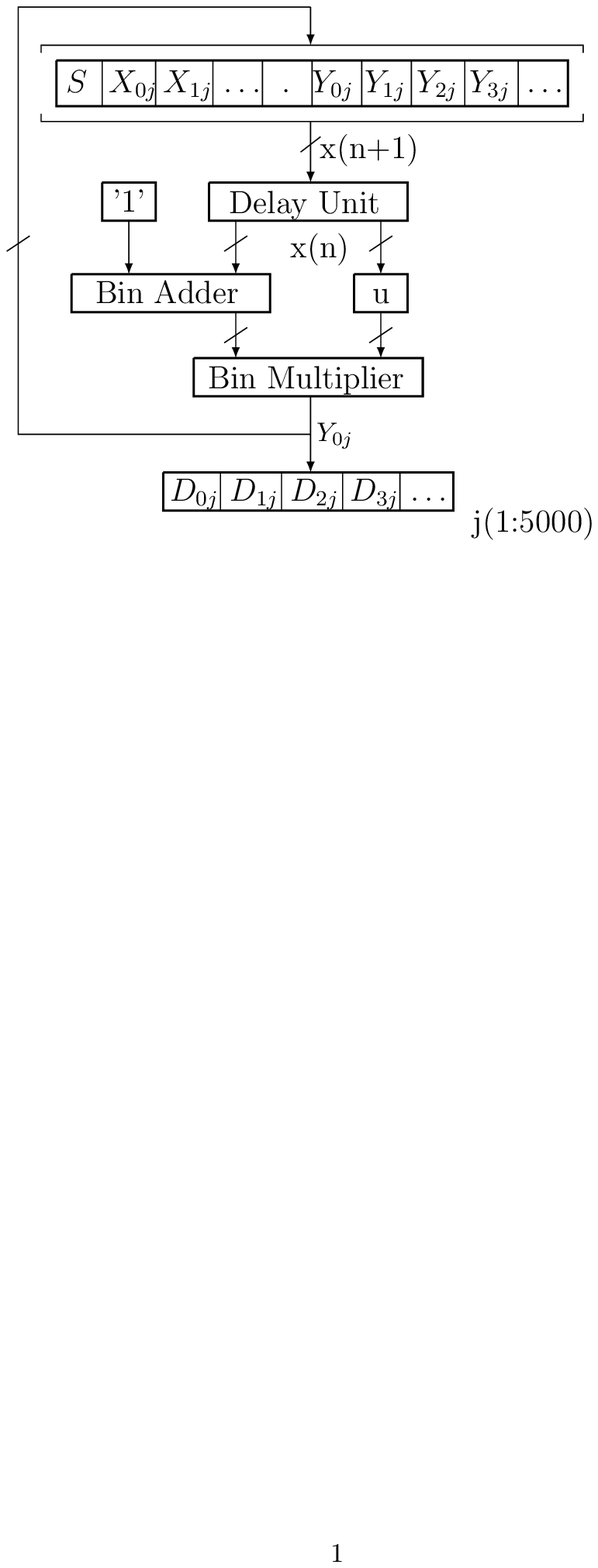}\\
 (b)\hspace{7cm}(c)
 \caption{(a) Creating random sequences from chaotic ones through a correction, (b) schematic diagram  of the chaotic generator with mixed-mode inputs, (c) example of a simplified digital fixed-point logistic map implementation.}
 \label{Fig2}
\end{figure*}

The choice of XOR operation of mixing two signals is due to two main reasons. First, the operation is realtively simple and does not require  complicated circuit realization. Secondly, the operation is widely used to mix the transmitted data with secret keys in typical secure electronic transmission implementations. Also, as demonstrated in this paper, the XOR operation significantly improves the quality of the obtained chaotic sequence - see sections 5 and 6 below. Also, the generator performs very well even when a small number of bit representation is used. Another important factor is that the quality analysis of our generator is done through a relativly simple, but reliable tool, namely the 0--1 test, that can be implemented in a real time monitoring system.

\section{Generation of chaotic bits}
The chaotic behavior of the logistic map occurs for a certain range of the parameter $\mu$. If the parameter value is chosen from that range, then for any initial condition $0<x(0)<1$, we obtain a sequence of real numbers in the interval $(0,1)$. Each of those real numbers is represented as a fixed point number. Those numbers are implemented in the digital structure of the logistic map shown in Fig. 1c. The $[S|X_{0j}|X_{1j}|...|\,\cdots \,|Y_{0j}|Y_{1j}|Y_{2j}|Y_{3j}|...]$ is a fixed point number representation. The $j$ index is the iteration number of the logistic map (or other implemented discrete map). The $S$ denotes the sign bit. The sign is the same in all iterations of the logistic map, becuase $x(n)\in (0,1)$ for all $n=0,1,2,\dots$, but it may change from iteration to iteration when other discrete maps are used. The $X_{ij}$ are the integer part bits, while the $Y_{ij}$ are the fractional part bits, respectively, in the $j$th iteration.        The output chaotic bits $\{D_i\}$ in Fig. 1c are obtained in each iteration by using a bit at a selected position  (fixed in all interations) in the sequence of consecutive  fixed point numbers.

Chaotic bits from the continuous system result from a threshold unit, for example in the form of a simple comparator \citep{2}. A synchronization unit is used to synchronize those bits with the occurrence of bits $\{D_i\}$. The discrete map is iterated to obtain $x(n+1)$ when a bit from the continuous chaotic system is received. The two independent chaotic sequences $\{D_i\}$ and $\{C_l\}$ are mixed by the XOR operation. As a result, a new sequence of bits $\{N_k\}$ is obtained. This sequence is next tested by the 0--1 test for chaos \citep{12,13}.

\section{The 0--1 test  for chaos} 
The 0--1 test is a relatively new tool used to test the presence of chaos in digital sequences when a mathematical model (system of equations) is not available. Thus, the 0--1 test is fundamentally different than the Lyapunov exponents method \citep{23}. The result of the 0--1 test has two forms: a single real number $K\in (0,1)$, and a two-dimensional graph with translation variables $p_c$ and $q_c$ \citep{14,15}.  For a chaotic sequence the number $K$ should be close to 1. Regular (nonchaotic) sequences result in numbers $K$ closer to 0. 
The values of $K$ can be computed by using two different methods: regression or correlation.

\begin{figure*}[!t]
 \centering
 \includegraphics[width=0.46\textwidth,height=2.0in]
 {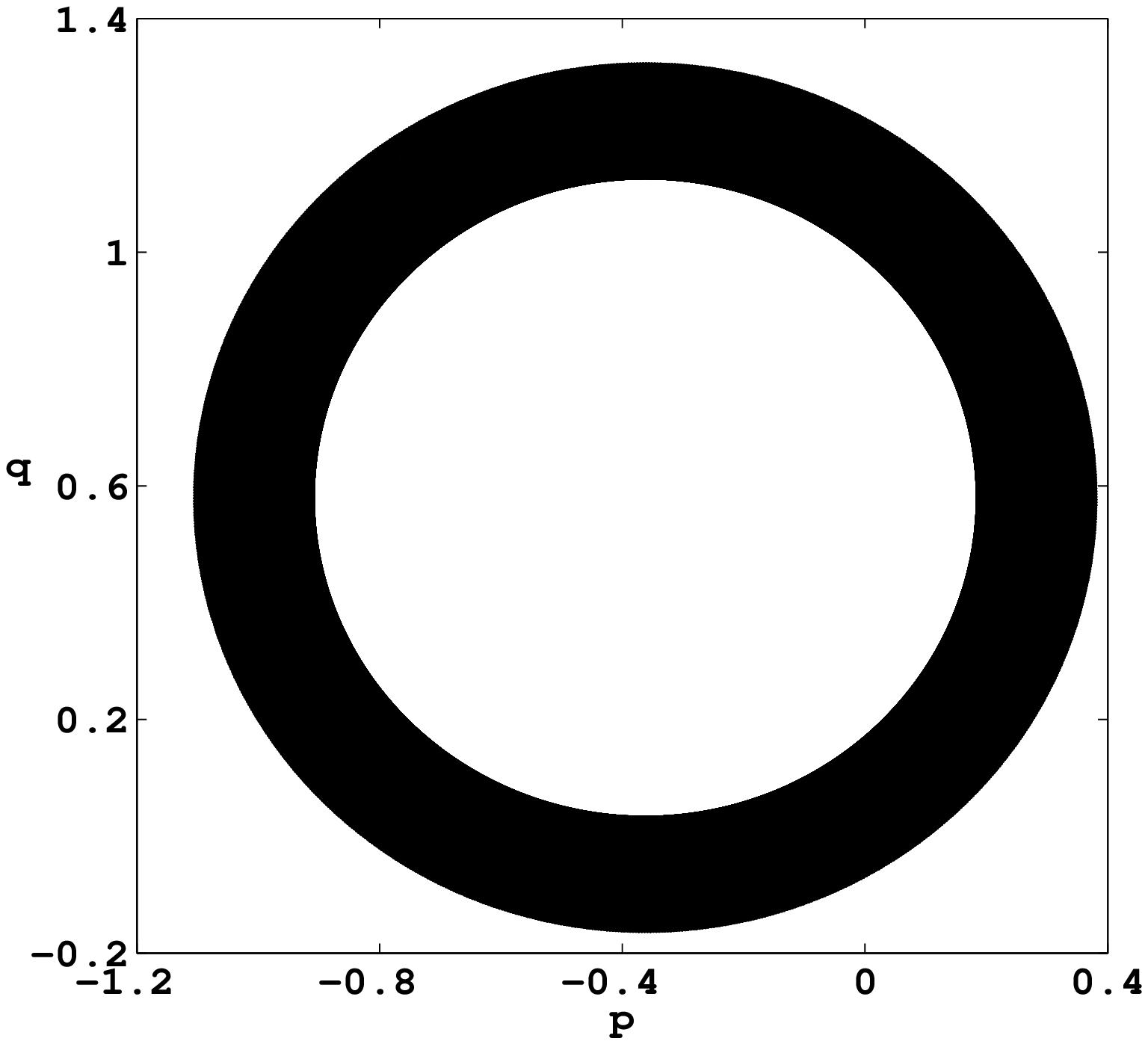}\hspace{0.5cm}
 \includegraphics[width=0.46\textwidth,height=2.0in]
 {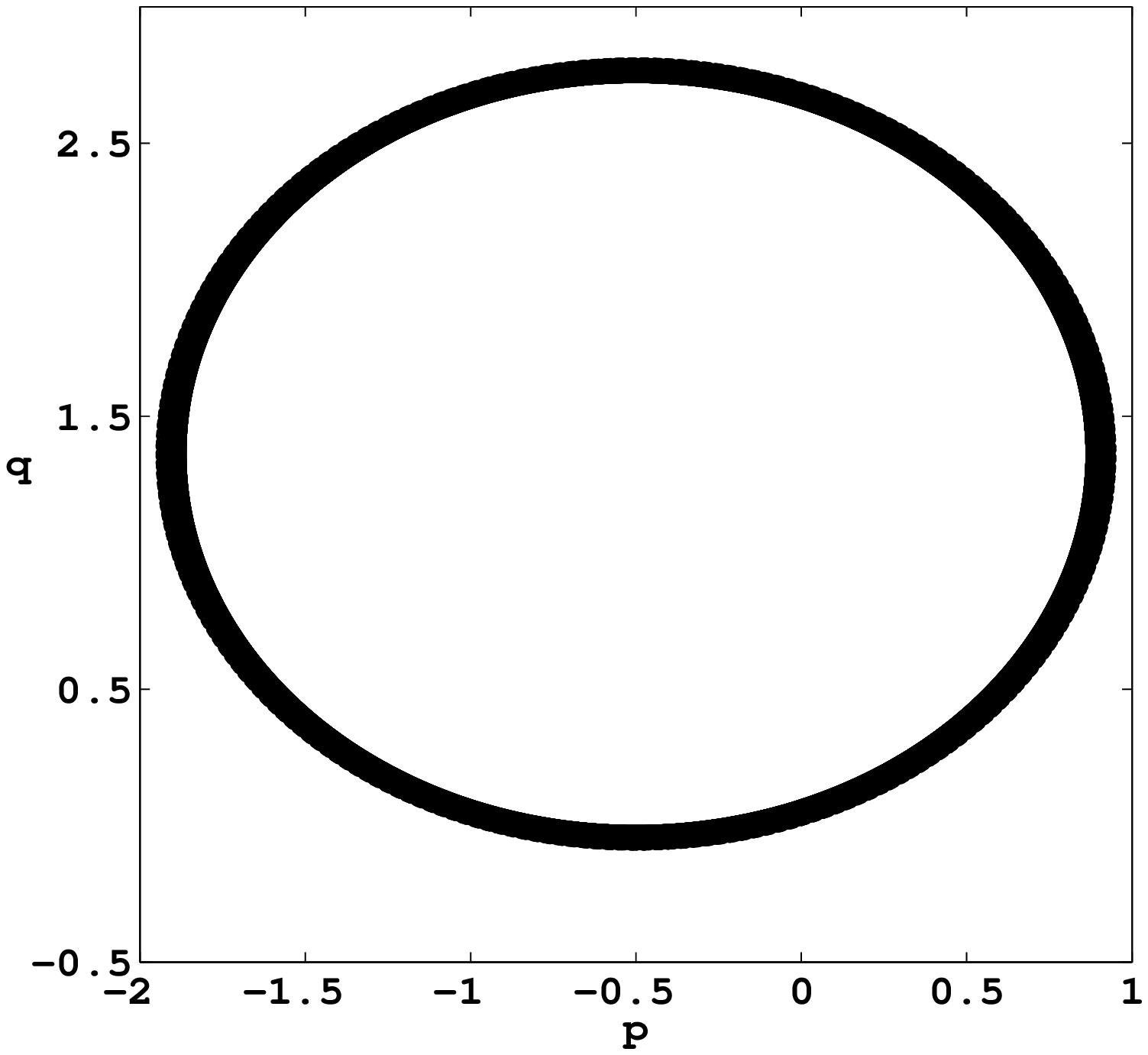}\\
 (a)\hspace{7cm}(b)\\
 \includegraphics[width=0.46\textwidth,height=2.0in]
 {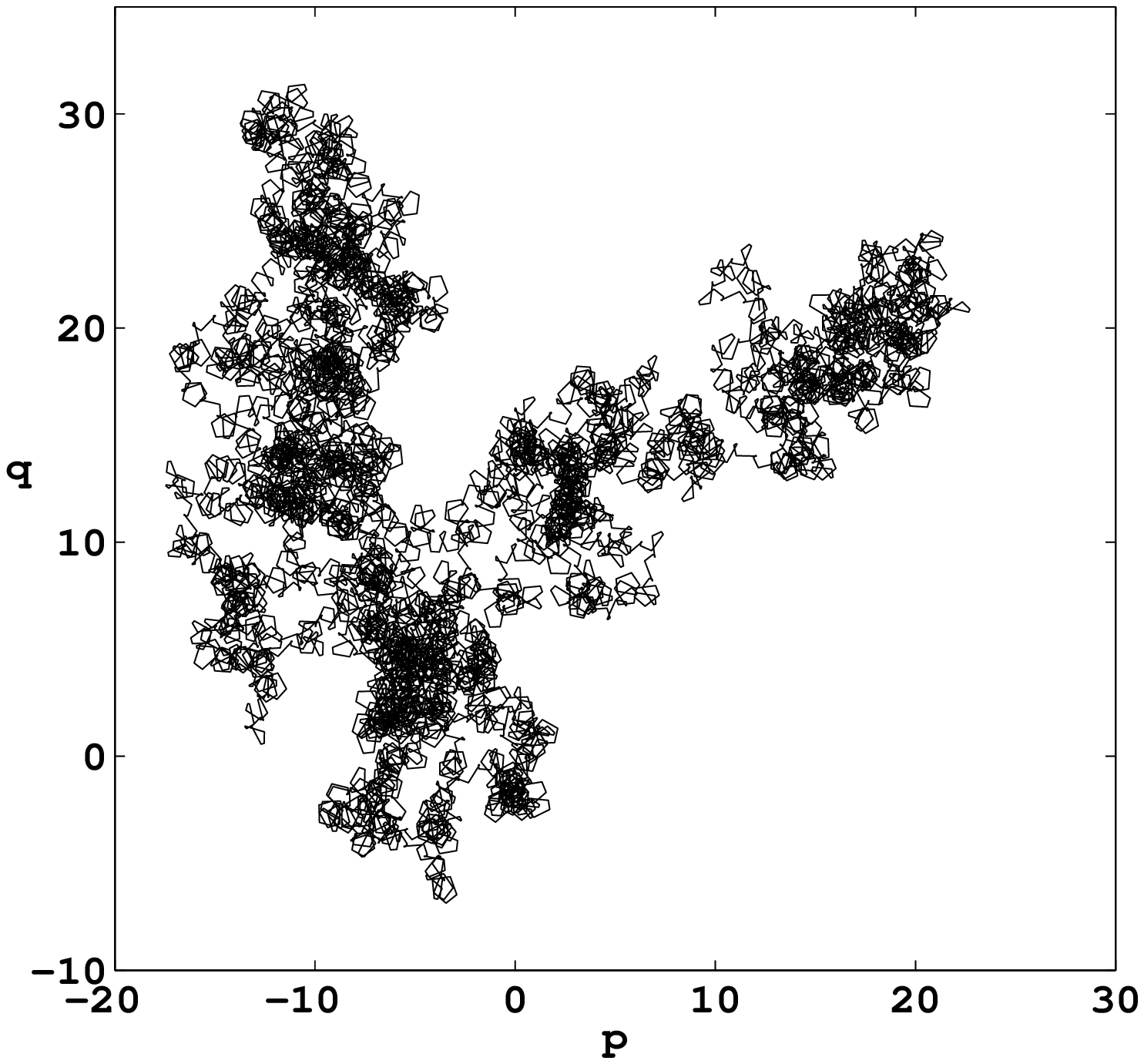}\hspace{0.5cm}
 \includegraphics[width=0.46\textwidth,height=2.0in]
 {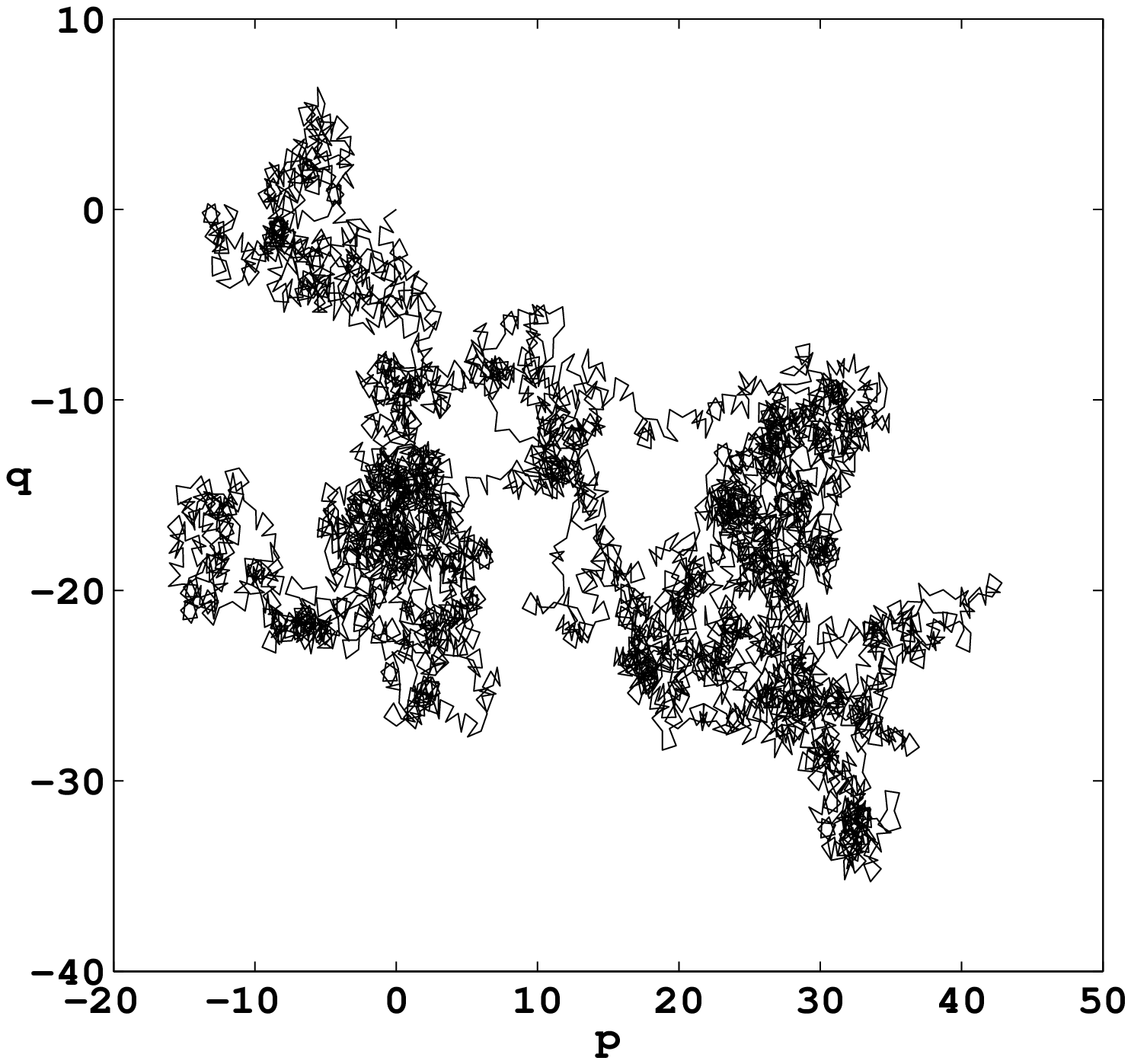}\\
 (c)\hspace{7cm}(d)
 \caption{Variables $q-p$ for: (a) logistic map only with $\mu=3.50$ yields $K=0.0015$, (b) sequence of $7th$ bits from logistic map with  $\mu=3.50$ yields $K=0.0025$, (c) logistic map only with  $\mu=3.99$ yields $K=0.9982$, and (d) sequence of $7th$ bits from logistic map with  $\mu=3.99$ yields $K=0.9980$.}
 \label{Fig4}
\end{figure*}

 For the sequence $\{N_k\}$, $k=0,1,\dots,\overline{N}-1$, the values $p_c$ and $q_c$  are computed by the following expressions for a randomly chosen real number number $c\in (0,\pi)$
\begin{equation}\label{eq1}
p_c(n)\!=\!\sum_{j=0}^n\! N_j cos[(j+1)c],\hspace{0.2cm}q_c(n)\!=\!\sum_{j=0}^n\! N_j sin[(j+1)c]
\end{equation}
for  $n=0,1,\dots,\overline{N}-1$. Then, the mean square displacement $M_c(n)$, $n=0,1,\dots ,n_{cut}$, of the variables $p_c(n)$ and $q_c(n)$ is computed with the recommended value $n_{cut}=(\overline{N}-1)/10$ \citep{13}
\begin{equation}\label{eq1n}
M_c(n)\!=\!\!
  \lim \limits_{\overline{N}\rightarrow \infty}\!\!\frac{1}{\overline{N}-1}\!\!\sum \limits_{j=0}^{\overline{N}-1}\!\!\left [p_c(j\!+\!n)\!-\!p_c(j)\right ]^2\!\!+\!\left [q_c(j\!+\!n)\!-\!q_c(j)\right ]^2\!\!\!.
\end{equation}
 Next, if the regression method is applied, then the asymptotic growth rate $K_c$ of the mean squared displacement is  computed as follows
\begin{equation}\label{eq1m}
K_c=\lim_{n\rightarrow \infty}\frac{log\,M_c(n)}{log\,n}.
\end{equation}
On the other hand, if the correlation method is applied, then two vectors $\Delta=(M_c(0),M_c(1),M_c(2),\dots,M_c(n_{cut}))$ and
  $\xi=(0,1,2,\dots,n_{cut})$ are created. The correlation coefficient $K_c$ is obtained as follows
\begin{equation}\label{eq56}
K_c=corr(\xi,\Delta)\equiv \frac{cov(\xi,\Delta)}{\sqrt{var(\xi)var(\Delta)}}
\end{equation}
where the \textit{cov} and \textit{var} stand for covariance and variance, respectively.

In both methods the above steps are repeated for $N_c$ values of $c$ chosen randomly  in the interval $(0,\pi)$. Again, papers \citep{12,13} recommend $N_c=100$. Finally, the median of the $N_c$ values of $K_c$ is the final number $K$. The $K\approx 1$ indicates a chaotic sequence of bits, while $K\approx 0$ indicates regular dynamics. More details about the 0--1 test, its properties,  reliability,  comparison with the FFT approach and application in the hardware trojan detection problems are considered in  \citep{15,28}.

Based on the sequences of $p_c$ and $q_c$ values one can create two-dimensional plots of $q_c$ versus $p_c$. For regular (nonchaotic) or weakly chaotic sequences (i.e. $K$ close to 0 or significantly lower than 1, respectively), the $q_c$  versus $p_c$ plot has a regular two-dimensional shape, while for chaotic sequences (i.e. $K$ close to 1) the shape is irregular. Numerical examples with both regular and irregular plots are given in the next section. In the examples we use the correlation method and drop the index $c$ in further analysis.

\section{Computational results}
We used sequences of $5000$ numbers in all our calculations. Also, $n_{cut}=10$ and $N_c=100$. Such values are suggested, for example, in \citep{14}.

\begin{figure*}[!t]
 \centering
 \includegraphics[width=0.46\textwidth,height=2.0in]
 {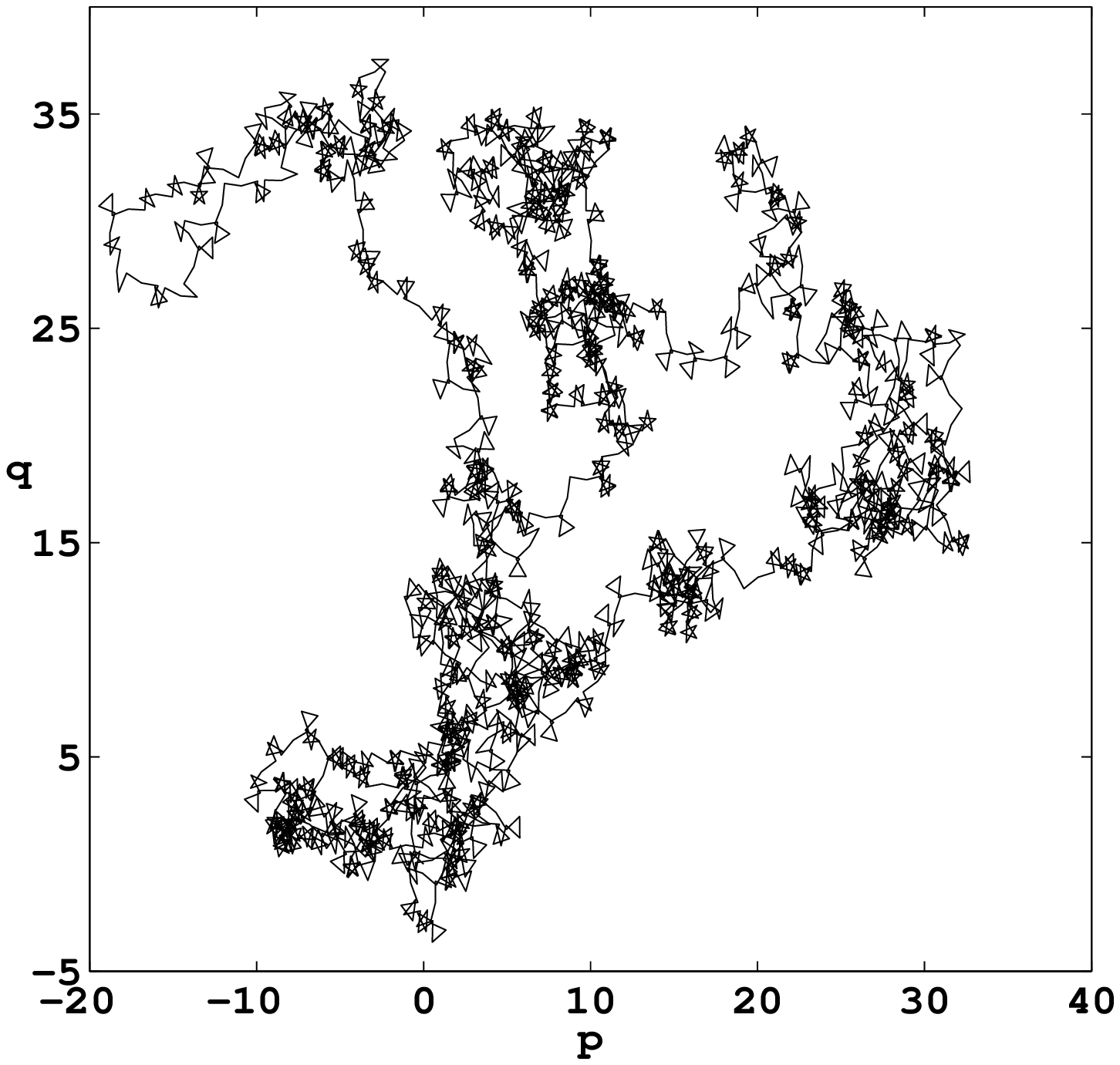}\hspace{0.5cm}
 \includegraphics[width=0.46\textwidth,height=2.0in]
 {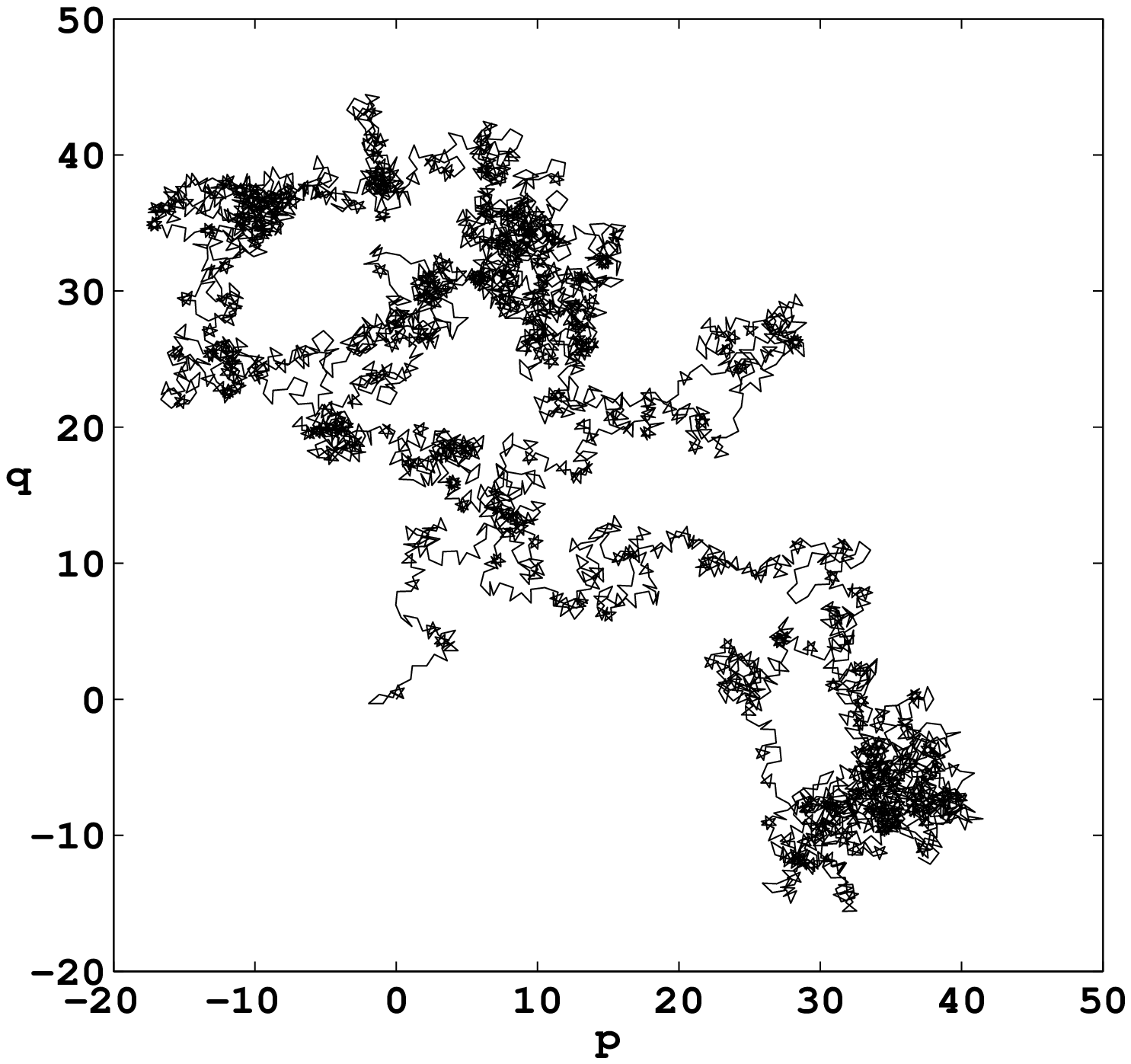}\\
 (a)\hspace{7cm}(b)\\
\includegraphics[width=0.46\textwidth,height=2.0in]
 {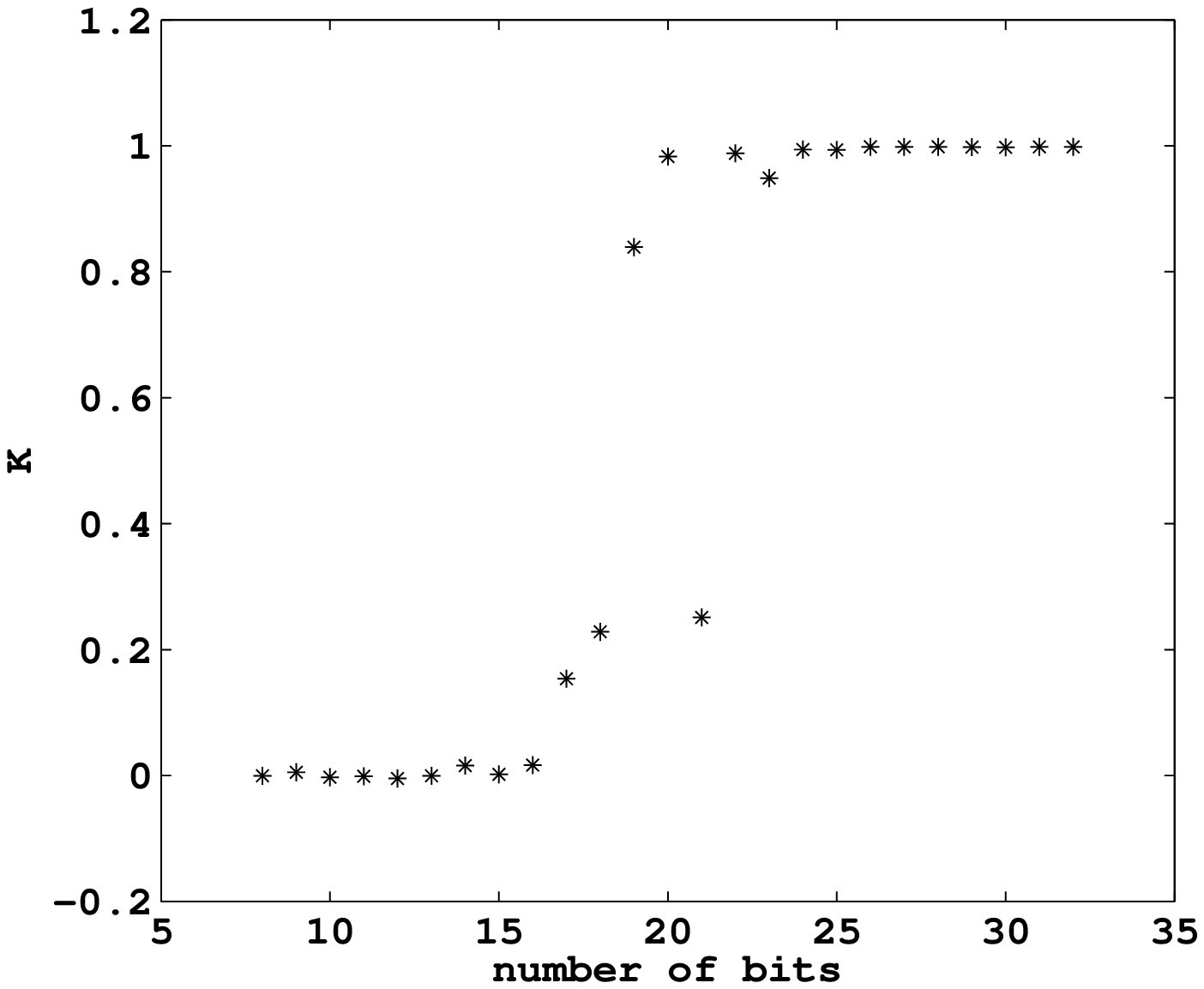}\hspace{0.5cm}
 \includegraphics[width=0.46\textwidth,height=2.0in]
 {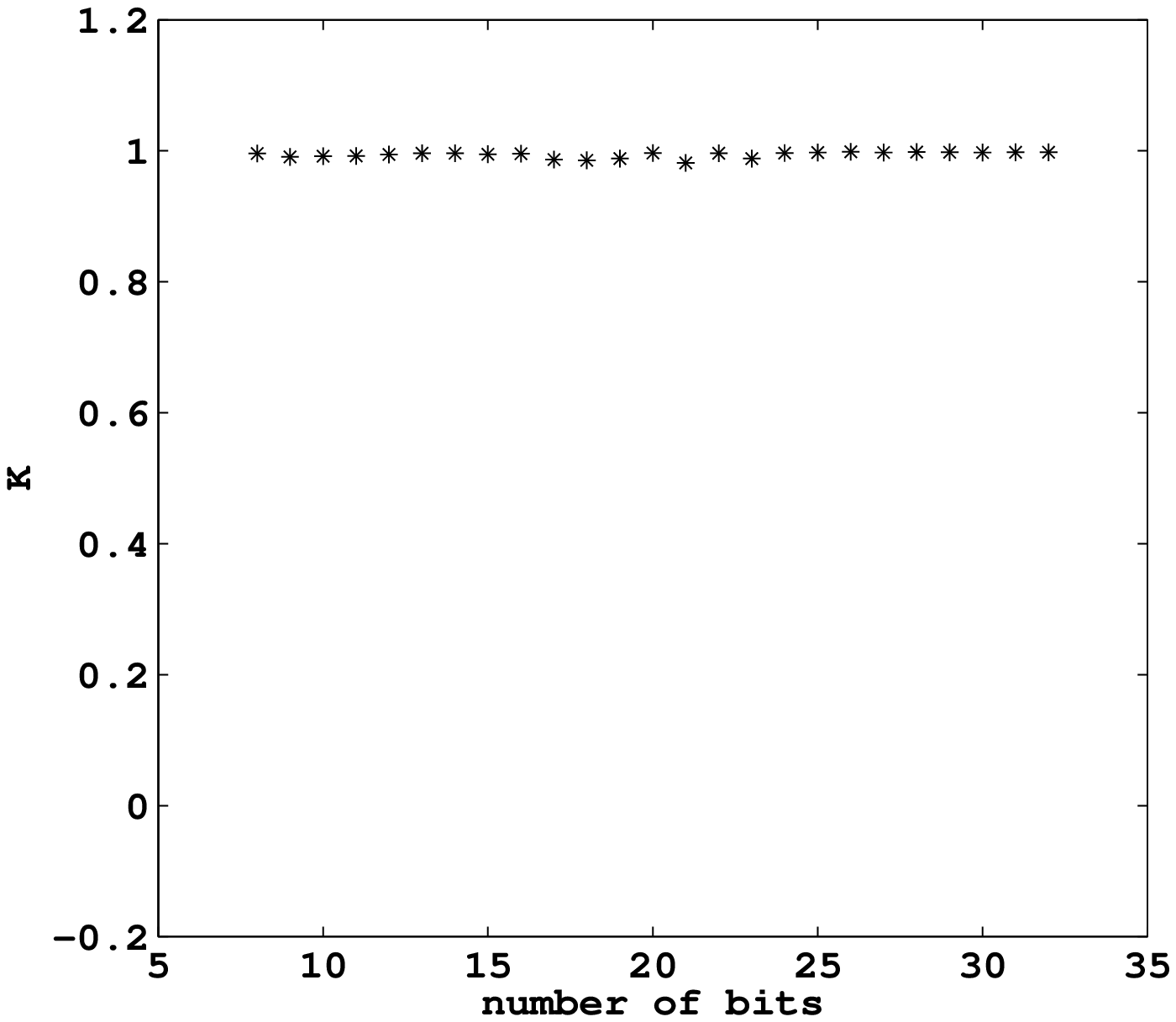}\\
(c)\hspace{7cm}(d)
\caption{(a) variables $q-p$ for a sequence of $7th$ bits of logistic map with $\mu=3.50$ XORed with chaotic
bits of the Chua (Matsumoto) circuit yield $K=0.9974$, (b) same as in part (a) but for $\mu=3.99$ results in $K=0.9984$, (c) 
 $K$ values for various number of bits between 8 and 32 (horizontal axis) for a sequence of $7th$ bits from the logistic map with $\mu=3.99$, (d)   same as in part (c) but the sequence of $7th$ bits from logistic map was XORed with the bits resulting from the continuous Chua (Matsumoto) circuit.}  
\label{Fig5}
\end{figure*}

First, we used the above approach to analyze the logistic map only, without a continuous chaotic system. For $\mu=3.50$  the following two cases were considered. First, we examined a sequence of real numbers from the logistic map. The $q-p$ plot is shown in Fig. 2a. Then, in the second case, the sequence of real numbers from the previous case was transformed to a respective fixed point representation. We selected a particular bit position and tested a sequence of bits from the consecutive fixed-point numbers, i.e.
 from each fixed-point number we selected a  single bit of $\left \{Y_{0j}\right \}$ in Fig. 1c (on the $7th$ position) and formed a
 sequence of bits taken from that selected position. Choosing the $7th$ position has no particular significance, and any other position could also be used. However, using a single bit position may have a significant impact and be an efficient implementation  method when hardware resources are limited, for example in FPGA devices. In this paper the logistic map is used in fixed point representation. Fig. 2b shows the $q-p$ plot obtained in the second case.  Since for $\mu=3.50$ the logistic map gives a nonchaotic signal, therefore, in both cases, the obtained $q-p$ graphs are of regular shape as shown in Figs. 2ab. An interesting result of our analysis is the fact that the nonchaotic nature of the sequence of real numbers (the first case above) is transformed into a nonchaotic nature of the sequence of bits $\{D_i\}$ (obtained from a selected single bit position as described in the second case above). The $K$ values in the two cases above are 0.0015 and 0.0025, respectively. These numbers clearly indicate a nonchaotic nature of the analyzed sequences.

The same two cases were analyzed for the logistic map with $\mu=3.99$ which result in chaotic behavior. The corresponding $q-p$ plots  are shown in Figs. 2cd, respectively. The irregular shapes in Figs. 2cd indicate a chaotic nature of the sequences. The corresponding $K$ values  are now very close to the number 1 and equal 0.9982 and 0.9980, respectively.

Further, Figs. 3ab show the $q-p$ results for the proposed mixed-mode generator with a continuous chaotic Chua generator XORed the  logistic map with a nonchaotic sequence ($\mu=3.50$, Fig. 3a) and also with a chaotic one ($\mu=3.99$, Fig. 3b). The $K$ values are equal 0.9974 and 0.9984, respectively.

The Chua (or other analog chaotic generator) can be realized as a hardware device (with elements $R$, $L$, $C$ and diodes together with op-amps) or as a software implementation - discretization of  solutions of  systems of differential equations \citep{17}.

Figs. 4cd show the results of applying the 0--1 test when a sequence of $7th$ bits were used for generated numbers with various bit lengths of sequence $Y$ (between 8 and 32). Logistic map with $\mu=3.99$ was used. It is clear from Fig. 4c that the chaotic sequences (the $K$ values close to 1) are obtained for numbers of length  22 and more bits. All numbers of 16 and fewer bits indicate regular (nonchaotic) sequences formed of the $7th$ bits. If the sequences of $7th$ bits (from logistic map) are XORed with the bits resulting from the continuous chaotic Chua (Matsumoto) circuit \citep{17}, then all the new sequences show chaotic nature with $K$ values close to 1, as shown in Fig. 4c. Moreover, we also fixed the length of numbers obtained from the logistic map (with $\mu=3.99$) to be 18 bits and, as shown in Fig. 4c, we obtained $K=0.2079$ for the sequence of $7th$ bits. The corresponding $q-p$ plot is shown in Fig. 4a. The $q-p$ plot is fairly regular (as expected for $K=0.2079$) and the sequence of $7th$ bits can, at best, be classified as \emph{weakly chaotic}. The same sequence, when XORed with a sequence obtained from continuous Chua (Matsumoto) circuit gives $K=0.9874$ and the $q-p$ plot is irregular, as
 shown in Fig. 4b.

\begin{figure*}[!t]
 \centering
 \includegraphics[width=0.46\textwidth,height=2.0in]
 {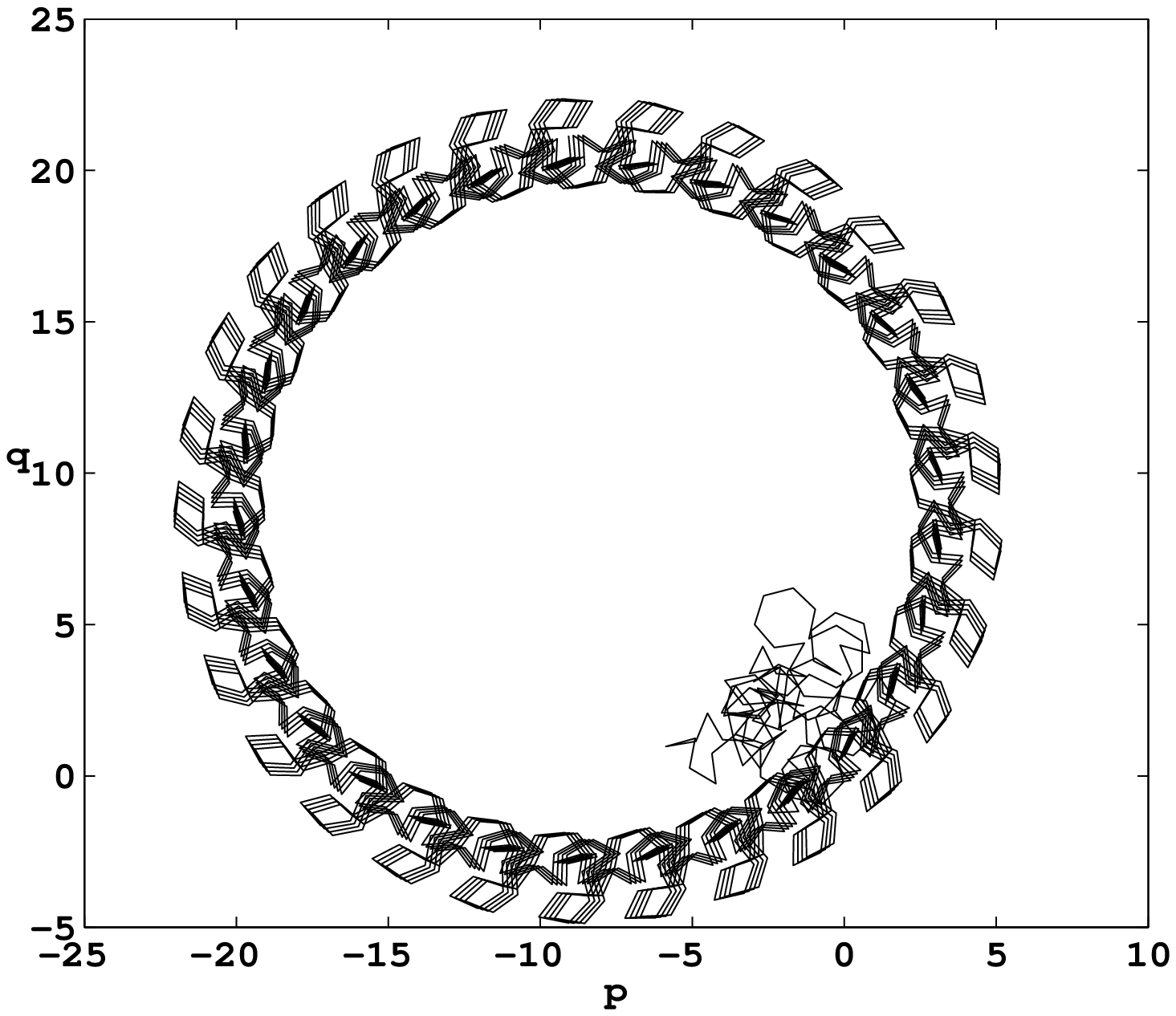}\hspace{0.5cm}
 \includegraphics[width=0.46\textwidth,height=2.0in]
 {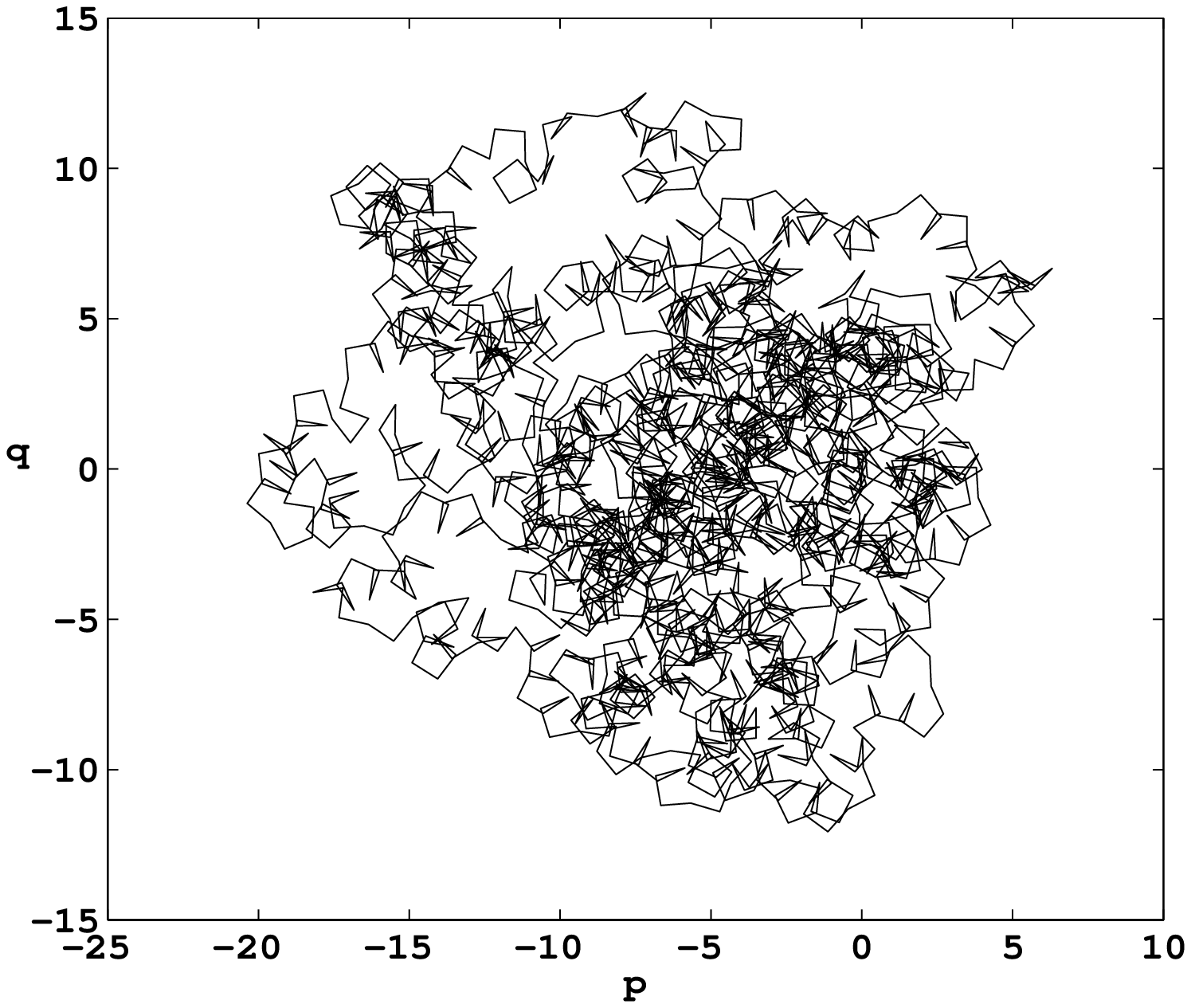}\\
 (a)\hspace{7cm}(b)\\
 \includegraphics[width=0.46\textwidth,height=2.0in]
 {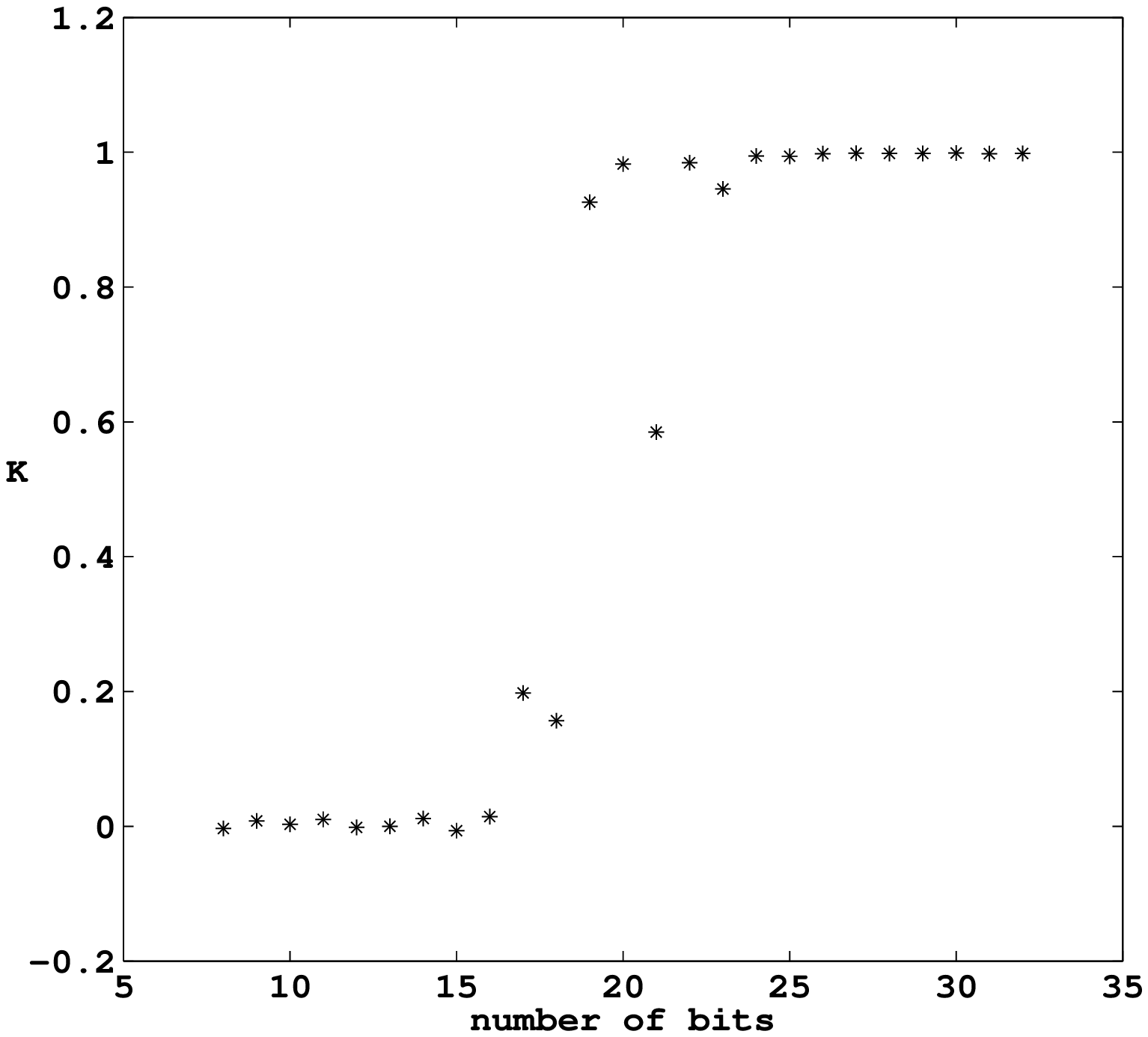}\hspace{0.5cm}
 \includegraphics[width=0.46\textwidth,height=2.0in]
 {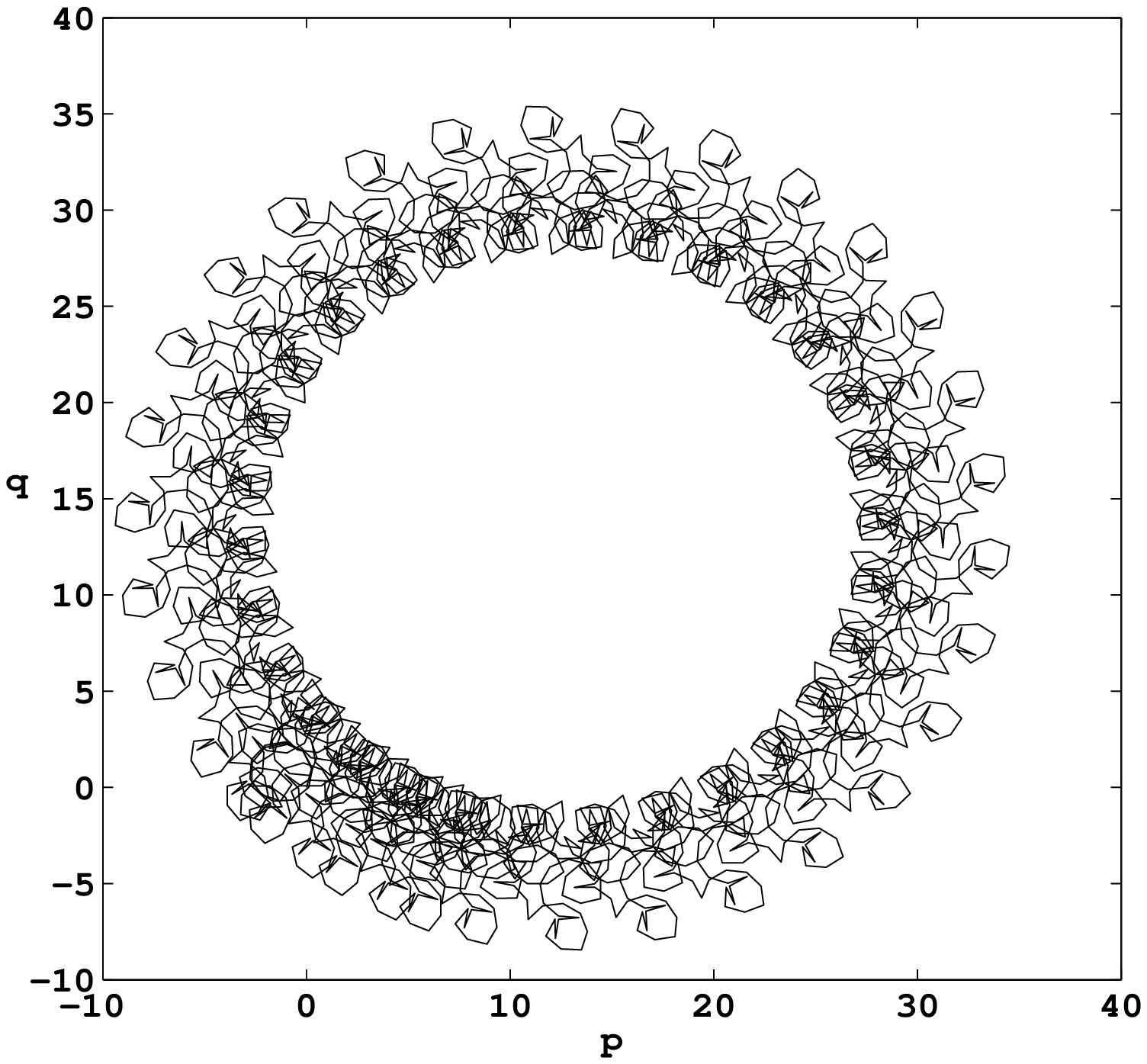}\\
 (c)\hspace{7cm}(d)
 \caption{(a) variables $q-p$ without  the XOR operation (chaotic bits from the Chua (Matsumoto) circuit). Logistic map with $\mu=3.99$ was used  for a sequence of $7th$ bits with numbers of length 18 bits ($K=0.2079$), (b) same as in (a) but with the XOR operation ($K=0.9874$), (c) $K$ values for various number of bits  of $\{Y_{0j}\}$ in Fig. 1c between 8 and 32 and the sequence of $7th$ bits from logistic map with $\mu=3.99$, (d) variables $q-p$ for a sequence of $7th$ bits when the total  length of each number is 21 bits. The $K=0.6038$ (Fig. 4c for the \emph{number of bits=21}).}
 \label{Fig7}
\end{figure*}

Fig. 4c shows a result similar to that of Fig. 3c, but it  illustrates another interesting fact about the 0--1 test. Notice that for a sequence with 21 bit representation we obtain $K=0.6038$, which is almost in the middle between the 0 and 1 values. The $q-p$ plot for this case is shown in Fig. 4d. The case can be classified as \emph{weak chaos} with a fairly regular plot $q-p$. 

We also examined our mixed-mode generator of chaotic bit sequences with various values of $\mu$ ($3.11\le \mu \le 3.99$) and XOR operation with other continuous chaotic systems, such as, the Chua system with Kennedy diode,   R\"{o}ssler and Lorenz chaotic continuous systems. The obtained $K$ values in all such cases were all greater than $0.99$ with differences in the third and fourth decimal digits.

Figs. 5ab show the bit sequences $\{D_i\}$, $\{C_l\}$ and $\{N_k\}$ for the Chua (Matsumoto) system with logistic map ($\mu=3.99$, Fig. 5a) and 
the Lorenz system with logistic map ($\mu=3.99$, Fig. 5b). The $\{N_k\}$ sequences yield $K=0.9981$ and $K=0.9983$ in Fig. 5a and Fig. 5b, respectively. Fig. 6 shows the result of using logistic map for $\mu=3.99$ and 12-bit precision of $\{Y_{0j}\}$. The value of $\mu=3.99$ should indicate a chaotic dynamics. However, due to a small number of bits used, we obtain a non-chaotic (or \emph{regular}) sequence $\{D_i\}$. The \emph{regular} sequence $\{D_i\}$ and chaotic sequence $\{C_l\}$ (Chua (Matsumoto) circuit) are XOR-ed to yield a new chaotic sequence $\{N_k\}$. The three sequences $\{D_i\}$, $\{C_l\}$ and $\{N_k\}$ in Fig. 6 are characterized by the values of $K$ equal 0.0011, 0.9980 and 0.9950, respectively.  Thus, Fig. 6 illustrates the case when one of the two input sources
is regular (non-chaotic sequence  $\{D_i\}$) and yet, thanks to the XOR operation with
another source sequence (with relatively small number of precision bits), we obtain
a strong chaotic sequence $\{N_k\}$.  Thus the analog part of the generator (Chua circuit) assures that the overall performance is firmly chaotic ($K=0.9950$), despite the fact that a  \emph{regular} sequence ($K=0.0011$) is obtained from the discrete part (logistic map) of the generator.

\section{Mixed-mode input chaotic generator: quality of randomness}

In the previous section we analyzed the chaotic dynamics of the generated bit sequences. 
In this section we evaluate whether or not the chaotic bit  sequencse have any 
features  of random processes. 
It should be underlined that the chaos phenomenon is not equivalent to a  random one. 
Some chaotic sequences show only selected random features, depending on the
length of the analyzed data. 
The goal now is to test if the mixed-mode generator's chaotic sequences exhibit similar random
features 
as the ideal random quantum generator does for sequences of short length. 

\subsection{The \textit{ent} tests}
The binary sequences obtained from the mixed-mode generator were tested by the \textit{ent} software and its various tests for randomness \citep{18}. A sequence of bits is first transformed  by \textit{ent} into ASCII characters.
 Then the sequence of ASCII characters  undergoes six independent tests, as follows \citep{19}.

\begin{itemize}
\item Entropy level test. For a  sequence of ASCII characters we
 obtain randomness if the entropy level is around the value of $8$. The lower the entropy level, the more likely it is to have
 a non-random sequence of ASCII characters. 
\item Compression test. 
Random sequences should have their compression levels close to $0 \%$. 
\item $\chi ^2$ (chi-square) test. Randomess is confirmed in this test if a sequence falls into the interval of $10 \%$ to $90 \%$. As explained in \citep{18}, such an interval is achieved primarily in the case of radioactive isotope decays. Also, the chi-square distribution is calculated for the stream of bytes in the sequence and
expressed as two values: an absolute number and a percentage which indicates how
frequently a truly random sequence would exceed the value calculated. For example,  the result of $(213.91;\,95\%)$ indicates a sequence with the $\chi ^2$  distribution  of $213.91$, and the sequence 
would exceed randomly that value $95\%$  of the times.
\item Arithmetic Mean Value (AMV) test with the output value  close to $127.5$ for random sequences. In this test all input bytes are summed up and divided by the total number of bytes.
\item Monte-Carlo $\pi$ (MC $\pi$) test indicating a random sequence if the result is a single percentage digit. For very long input streams this value will be close to 0, meaning an accurate approximation of the number $\pi$ \citep{18}. 
\item Serial Correlation Coefficient (SCC) test yielding the number close to $0.0$ for  random sequences. This test checks dependece of each byte  on a previous one. If  there is no dependence between bytes, then the SCC value is close to $0.0\,.$
\end{itemize}

The details of the tests can be found in \citep{18,19}.
All types of chaotic bit sequences discussed in this paper were tested by the above six tests. We list these sequences in Table 1 and mark them as sequences $s1, s2,\dots,s9$. The most important test for randomness is, in our opinion, the entropy test. Therefore, as a reference sequence $s1$ we selected a sequence obtained from a commercially available quantum  generator Quantis manufactured by the Swiss firm \citep{10}. We used the model USB-4M with the serial number $163109A410$ supported by Quantique's official ``True Quantum Randomness Certificate" \citep{20}.
 The sequence $s1$, as the one with excellent parameters (perhaps with the exception of the $\chi ^2$ value), is a reference sequence to which we relate  all other chaotic sequences, including those obtained with the XOR operation in our mixed-mode generator.

\begin{table}[h!]
\caption{\label{label}Sources of bits sequences tested by \emph{ent}}
\begin{tabular}{cl}
\hline \hline
sn&Source\\
\hline
{\bf s1}  &{\bf  Quantis' reference sequence of high entropy \citep{10}}  \\
s2  & Lorenz chaotic  bits  \\
s3  & Chua  chaotic bits   \\
s4  & Logistic chaotic bits (32 bit sequence $\{Y_{7j}\}$, see Fig.1c) \\
s5  & Logistic chaotic bits (10 bit sequence $\{Y_{7j}\}$, see Fig.1c)  \\
\hline
\textcolor{blue}{s6}  &\textcolor{blue}{sequence s2 XOR sequence s4}\\
\textcolor{blue}{s7} & \textcolor{blue}{sequence s3 XOR sequence s4}\\
\textcolor{magenta}{s8} &\textcolor{magenta}{sequence s2 XOR sequence s5} \\
\textcolor{magenta}{s9}  &\textcolor{magenta}{sequence s3 XOR sequence s5} \\
\hline \hline
\end{tabular}
\end{table}

\subsection{Analysis of  test results}

\begin{figure*}[b!]
 \centering
 \includegraphics[width=0.8\textwidth,height=1.80in]
 {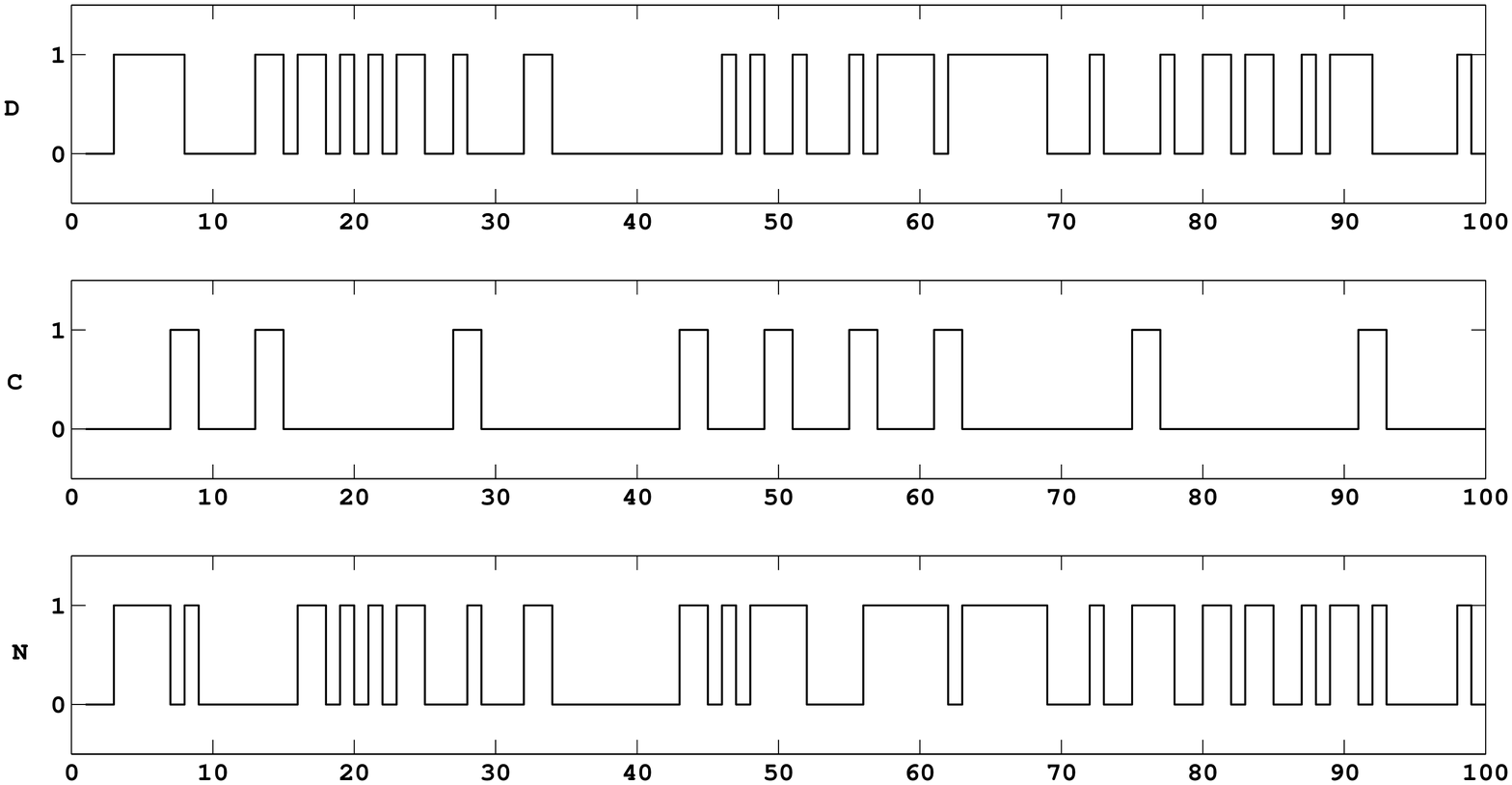}\\
 (a)\vspace{0.4cm}\\
 \includegraphics[width=0.8\textwidth,height=1.8in]
 {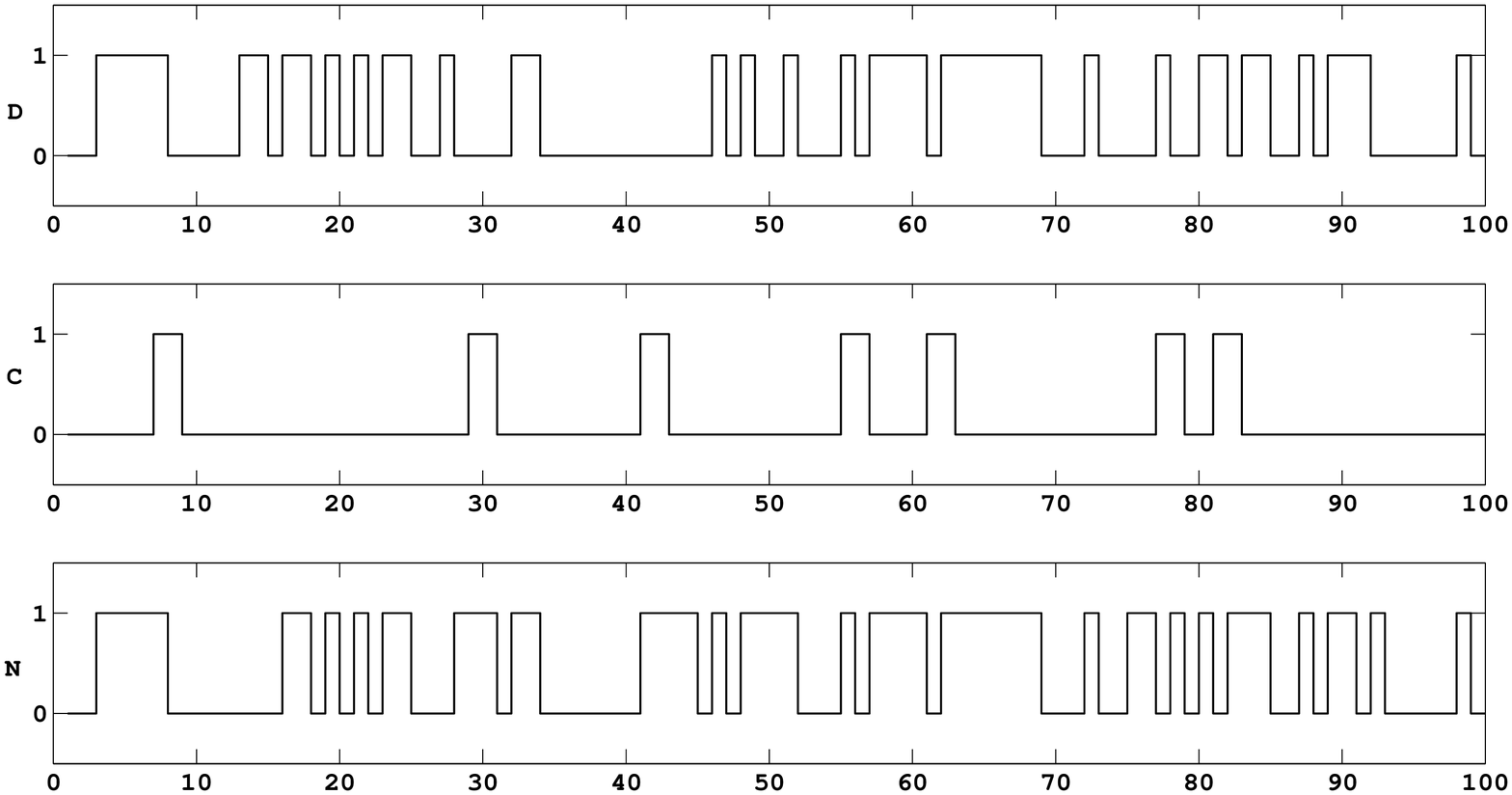}\\
(b)
 \caption{(a) sequences $\{D_i\}$, $\{C_l\}$ and $\{N_k\}$ obtained from the Chua (Matsumoto) system and logistic map with $\mu=3.99$, (b) same as in (a) but for Lorenz system and logistic map.}
 \label{fig9}
\end{figure*}

\begin{figure*}[!h]
 \centering
 \includegraphics[width=0.8\textwidth,height=1.8in]
 {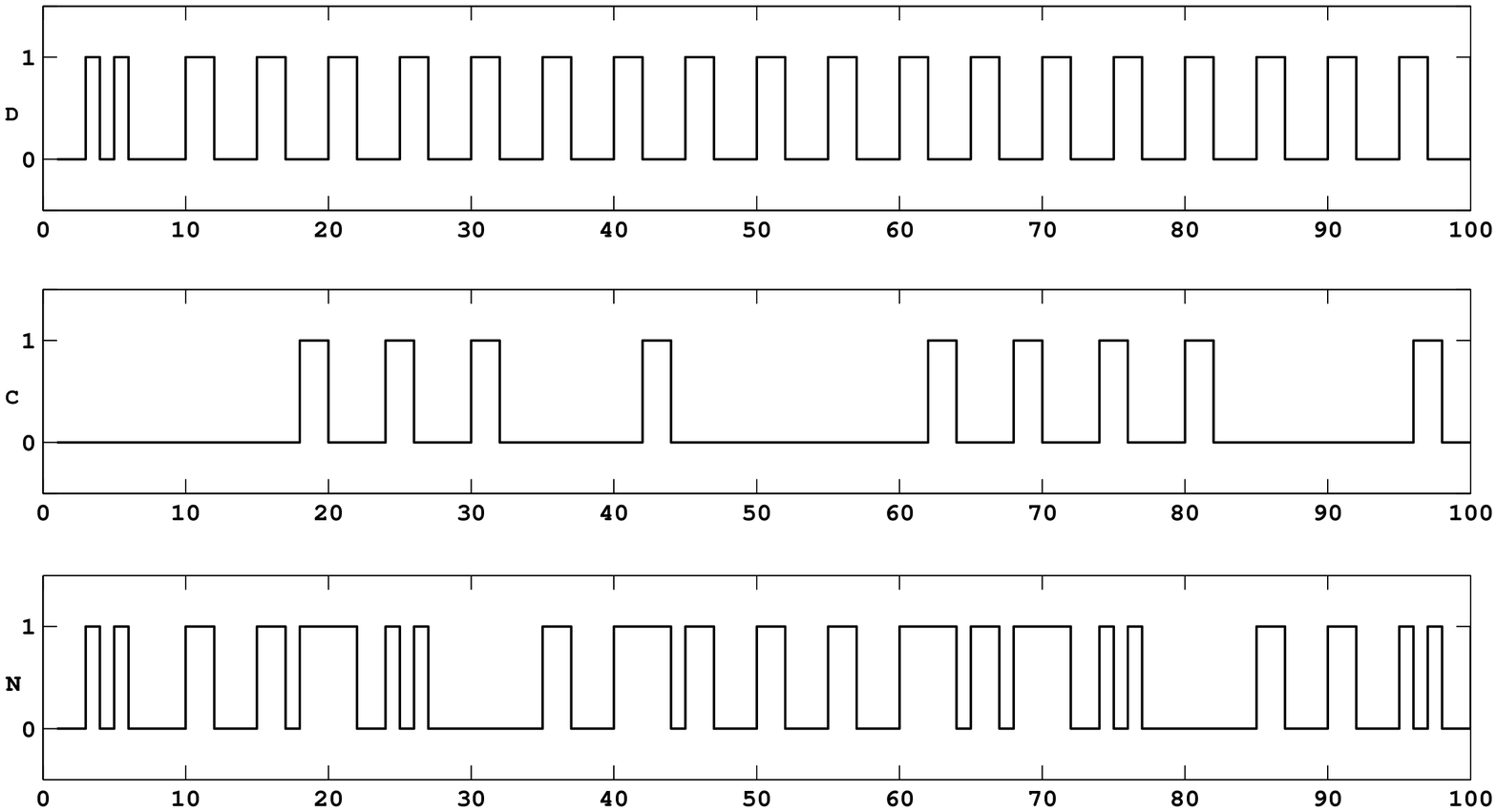}
\caption{Sequences as in Fig. 5 with a relatively small number of precision bits resulting in a nonchaotic sequence  
$\{D_i\}$.}
 \label{fig11}
\end{figure*}

Fig. 7  shows examples of scatter plots of $s1$--$s9$, each consisting of $10,000$ bits stacked in rows of $100$ bits. Visual and rather subjective observation of the nine sequences in Fig. 7  indicates that the scatter plots of sequences $s6$ (= $s2$ XOR $s4$) and $s7$ (= $s3$ XOR $s4$) are very close to the \emph{ideal} sequence $s1$. The scatter plot obtained from the logistic equation with 32 bit respresentation (sequence $s4$) also seems to be of good quality, but the same logistic equation yields much worse scatter plot if 10 bits are used only (sequence $s5$). The scatter plots of sequences $s2$ and $s3$ seem to be of low quality when compared to the scatter plot of $s1$. Sequences $s2$ and $s3$ and their scatter plots depend heavily on the threshold value
of the Threshold Unit in Fig. 1b as well as on the method of registering of chaotic bits and the frequency of the internal clock. Those parameters should be adjusted experimentally for various chaotic systems.  The scatter plot of  $s9$ (= $s3$ XOR $s5$) confirms, at least visually,
that the low quality sequence $s5$ has been improved after the XOR operation. The above visual observations of the scatter plots in Fig. 7  are, to large extend, confirmed by the results of the six \emph{ent} tests, which are analyzed below.
The results shown in Table 2 for sequences $s2$ (Lorenz binary chaotic sequence) and $s3$ (Chua binary chaotic sequence) indicate 
 relatively low entropy level, much lower than the desired value of $8$. This is caused by the bias phenomenon - long intervals of the
 same bits. Thus, neither $s2$ nor $s3$ can be considered as a  random sequence with high levels of entropy and the von
 Neumann correction is certainly recommended to those individual sequences. The logistic chaotic bit sequence $s4$ performs
 surprisingly well with a high entropy level (the length of data in $\left \{Y_j\right \}$ in Fig. 1c is 32 bits), and much poorer if the length 
is only 10 bits (sequence $s5$ in Table 2). The $s4$ sequence is generated by the system in Fig. 1c. There is a danger of having a
 repeated, identical sequences $s4$, if the system uses the same initial seed value. This can happen when an unathorized (hacking)
 event occurs.

\begin{table*}[h!]
\small
\caption{Results of 6 \emph{ent} tests performed on sequences $s1$--$s9$}
\begin{tabular}{ccccccc}
\hline \hline
sn&Entropy & Comp. \%&$\chi ^2$ value; \%&AMV &MC $\pi$ \%&SCC\\
\hline 
{\bf s1} &{\bf 7.869559}  & {\bf 1}   &{\bf 213.91; 95}    &{\bf 127.4415}   & {\bf 1.07}   &{\bf 0.005292}\\
s2 &3.884870  &51  &36200.14; 0.01 & 62.4149   &21.87   &0.119915\\
s3 &1.829749  &77  &162993.87; 0.01 & 39.1188   &27.32   &0.111901\\
s4 &7.850575  & 1   &263.07; 50    &131.5946  & 4.81    &0.051919\\
s5 &2.845031  &64  &44492.49; 0.01   &181.7886  &44.22   &0.037565\\
\hline
\textcolor{blue}{s6} &\textcolor{blue}{7.851689}  &\textcolor{blue}{1} &\textcolor{blue}{251.59; 50} &\textcolor{blue}{129.7392} &\textcolor{blue}{2.06} &\textcolor{blue}{0.013551}\\
\textcolor{blue}{s7} &\textcolor{blue}{7.837249}  &\textcolor{blue}{2} &\textcolor{blue}{270.84; 25}&\textcolor{blue}{131.6376}  & \textcolor{blue}{2.67} &\textcolor{blue}{0.043232}\\
\textcolor{magenta}{s8} &\textcolor{magenta}{6.266494}&\textcolor{magenta}{21}  &\textcolor{magenta}{5044.32; 0.01}&\textcolor{magenta}{155.7672}  &\textcolor{magenta}{16.14}   &\textcolor{magenta}{0.102282}\\
\textcolor{magenta}{s9} &\textcolor{magenta}{4.512645}  &\textcolor{magenta}{43}  &\textcolor{magenta}{22535.47; 0.01}&\textcolor{magenta}{166.5360}&\textcolor{magenta}{32.67}&\textcolor{magenta}{0.073107}\\
\hline \hline
\end{tabular}
\normalsize
\end{table*}

Also, having a finite number of bits available to represent initial condition (seed value), there is a  danger of
 inserting the same seed value after many repeated cycles of using a single input logistic-based generator.  Such a  generator will output sequences that had already been generated
 before. In order to secure a much wider diversity in creating chaotic sequences with excellent randomness features, we examined
 how the sequences obtained from our mixed-mode generator, perform in the six tests. The sequences $s6$ and $s7$ (obtained by
 the XOR operation in our generator) show very good test results, certainly comparable with the results for sequence $s1$. One may
 argue that there is no significant improvement of the test result between the $s4$ and $s6$ (or $s7$) sequences. However, even in
 this case, using the mixed-mode generator we have the comfort  of not having identical sequences that may be obtained when only a logstistic map is used with the same initial seed values. Clearly, if we use the logistic map only with data of $10$ bits (sequence
 $s5$), then such a single source generator fails most of the tests and $s5$ is of a low entropy level. XORing either $s2$ with $s5$ or
 $s3$ with $s5$ in our mixed-mode generator, creates sequences $s8$ or $s9$, respectively. These sequences have better test
 results than $s5$ alone with the entropy level increased two-fold. There are also improved  compression levels, $AMV$ values and
 MC $\pi$ percentages for $s8$ and $s9$ sequences when compared to the $s5$ sequence. Our results of the MC $\pi$ test are at the desired 1-2$\%$ level for $s6$ and $s7$. Such results are typically obtained with much longer sequences \citep{18}.
The results of the $\chi ^2$ test for $s6$ and $s7$ are also much better than those of $s2$ and $s3$ (and even of $s1$). As described on the \citep{21} site, the $\chi ^2$ values of weakly random sequences are large while small for random sequences. This is clearly seen from Table 2. Those results are also confirmed by the lower compression levels for $s6$ and $s7$, which are in the range of 1-2$\%$, comparable with the compression level for $s1$. Also, let us not forget that {\bf none} of
 the sequences $s6$, $s7$, $s8$, and $s9$ undergoes the von Neumann correction. Overall, these sequences have good 
 characteristics of random sequences, often comparable with those obtained from the professional quantum random number generators, such as the one used in this paper to generate sequence $s1$. 

\begin{figure*}[!t]
 \centering
 \includegraphics[width=0.99\textwidth]
 {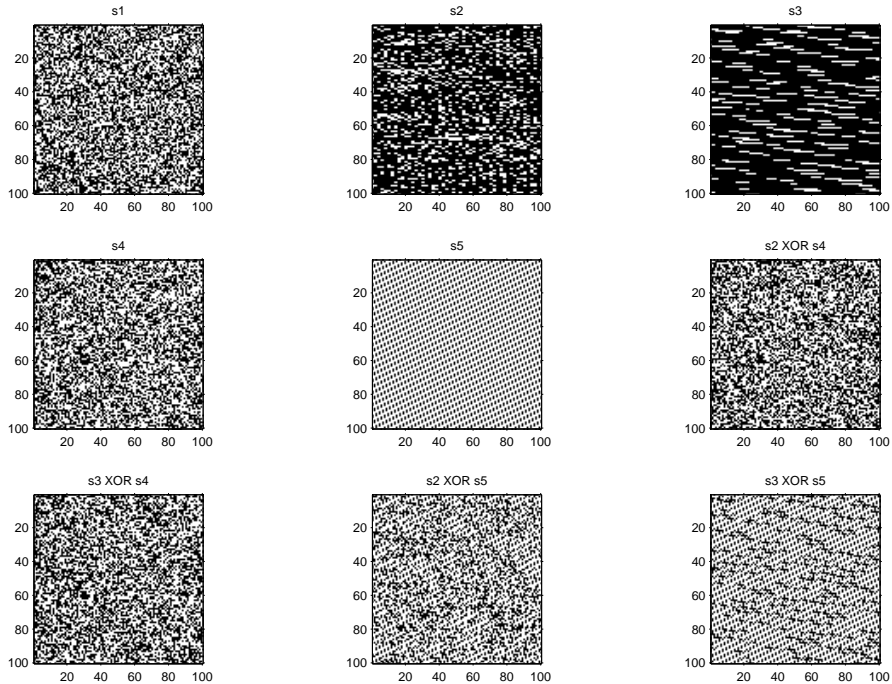}
 \caption{Scatter plots of sequences $s1$-$s9$, each with $10,000$ bits.}
 \label{fig12}
\end{figure*}

 \section{Conclusion}
 At the present time the methods of designing of  chaotic bit generators lack the  protection against a potential of synchronization,  and problems of finite length of bits in a number representation. Also, a possibilty of reusing of the same seed values (initial cnditions), obtained from chaotic generators to initialize pseudo-random algorithms, has not been properly addressed and examined yet \citep{6}-\citep{8}. To deal with those serious problems, 
we have proposed a new mixed-mode generator of chaotic bit sequences with an increased level of chaotic behavior (as evidenced by the increased parameter $K$ in the 0--1 test). The improvement is due to mixing two independent chaotic inputs (continuous and discrete). The computational results confirm the \emph{strong} chaotic nature of the generator's output as evidenced by the $K$ values and irregularity of the $q-p$ shapes in the 0--1 test. The two chaotic sequences that we mix through the XOR unit are also checked individually for their quality. Our generator exhibits \textit{some} features of randomness without the von Neumann correction that is needed in most other chaotic based random bit generators. Also, it is worth mentioning that for shorter binary sequences, such as considered in this paper, the NIST statistical tests are not applicable. Therefore, we use the \textit{ent} suite to examine the generator's performance and compare it with the performance of a 
truly random quantum reference generator of Quantis \citep{10,20}. Our chaotic generator can be assembled as a classical electronic circuit (continuous chaotic system) hooked up to a FPGA yielding a discrete chaotic system. In such a case, to model the whole generator one can use hardware description languages suitable to analyze mixed systems (continuous and discrete) \citep{25}-\citep{26}. Finally, our approach of taking the 7th bit in consecutive numbers is an efficient way to be implemented in FPGA devices when the  hardware resources are very limited.

The future  monitoring system built to check the real-time performance of the mixed-mode generator can utilize the graphical result in the form of a 2D $q-p$ plot, supplemented by the easy to interpret number $K$. A decrease of $K$ from, say, $0.9950$, through $0.8500$ and further to, say, $0.7200$ would indicate worsening the quality of chaotic output and could indicate improper work of the generator, due to a failure of electronic circuits generating the two input sequences. Another reason of decreasing the values of $K$ could be unauthorized hardware modifications defined 
as hardware attacts or hardware trojans. They occur in the forms of
parameter or circuit diagram changes, and became lately a \textit{hot} research topic  \citep{26a,27}.


\begin{thebibliography}{99}

\bibitem{1} Kanaso, A.,  Smaoui, N.: `Logistic chaotic  for binary numbers generations', \textit{Chaos Solitions Fractals}, 2009, \textbf{40}, pp. 2557--2568

\bibitem{2} Yalcin, M. E., Suykens, J. A. K.,  Vandewalle, J.: `True random bit generation from a double-scroll attractor',  \textit{IEEE Trans. Circuits  Syst.--I: Regular Papers}, 2004, \textbf{51}, pp. 1394--1404

\bibitem{3} Posadas-Castillo, C., Cruz-Hernandez, C., Lopez-Gutierrez, R. M.:  `Experimental realization of synchronization in complex networks with Chua's circuits like nodes', \textit{Chaos Solitons Fractals}, 2009, \textbf{ 40}, pp. 1963--1975 

\bibitem{4} Wu, X., Chen, G., Cai, J.: `Chaos synchronization of the master-slave generalized Lorenz systems via linear state error feedback control', \textit{Physica D}, 2007, \textbf{229}, pp. 52--80

\bibitem{4a} Marszalek, W., Trzaska, Z. W.: `Memristive circuits with steady-state mixed-mode oscillations',  \textit{Electr. Lett.}, 2014, \textbf{50},  pp.  1275-1277

\bibitem{4b} Marszalek, W., Trzaska, Z. W.:  `Mixed-mode oscillations and chaotic solutions of jerk (Newtonian) equations',  \textit{J.  Comp. Appl.  Math.}, 2014, \textbf{262}, pp. 373-383


\bibitem{5} Yang, Y., Zhu, F.: `Synchronization for chaotic systems and chaos-based secure communications via both reduced-order and step-by-step sliding mode observers', \textit{Commun. Nonl. Sci. Numer. Simul.}, 2013, \textbf{18},  pp. 926--937

\bibitem{6} Persohn, K. J.,  Povinelli, R. J.: `Analyzing logistic map pseudorandom number generators for periodicity inducted by finite precision floating-point representation', \textit{Chaos Solitions Fractals}, 2012, \textbf{45}, pp. 238--244

\bibitem{7} Alvarez, G.,  Li, S.: `Breaking an encrypted scheme based on chaotic baker map',  \textit{Phys. Lett. A},  2006, 
\textbf{352}, pp. 78--82

\bibitem{8} \"{O}zkaynaka, F.,  Yavuzb, S.: `Security problems for a pseudorandom sequence generator based on the Chen chaotic system', \textit{Computer Phys. Commer.}, 2013, \textbf{184}, pp. 2178--2181

\bibitem{9}  Von Neumann, J.: `Various techniques used in
connection with random digits', \textit{Appl. Math. 
Series}, 1951, \textbf{12}, pp. 36--38

\bibitem{10} `ID Quantique SA White Paper Version 3.0. Random number generation using quantum physics', http://www.idquantique.com, accessed July 2017

\bibitem{12} Gottwald, G. A.,  Melbourne, I.: `A new test for chaos in deterministic systems',  \textit{Proc.  Royal Soc. London}, 2003, \textbf{460}, pp. 603--611

\bibitem{13} Gottwald, G. A.,  Melbourne, I.: `Testing for chaos in deterministic systems with noise', \textit{Physica D}, 2005, \textbf{212}, pp. 100--110

\bibitem{23} Marszalek, W., Trzaska, Z.:  `Mixed-mode oscillations in a modified Chua's circuit', \textit{Circuits Syst. Signal Proc.},  2010, \textbf{29}, pp. 1075--1087

\bibitem{14} Gottwald, G. A., Melbourne, I.:  `On the implementation of the 0--1 test for chaos', \textit{ SIAM J. Appl. Dyn. Syst.}, 2009, \textbf{8}, pp. 129--145

\bibitem{15} Melosik, M.,  Marszalek, W.: `On the 0-1 test for chaos in continuous systems', \textit{Bull. Polish Acad. Sci., Tech. Sci.},  2016, \textbf{64}, pp. 521--528

\bibitem{28} Melosik, M.,  Marszalek, W.: `Using the 0-1 test for chaos to detect
hardware trojans in chaotic bit generators',  \textit{Electr. Lett.}, 2016, \textbf{52}. pp.  919-920

\bibitem{17} Matsumoto, T.,   Chua, L. O.,  Komuro, M.: `The double scroll',  \textit{IEEE Trans. Circuits Syst.},  1985, \textbf{CAS-32}, pp. 798--818

\bibitem{18} Walker, J.: `ENT. A psudorandom number sequence test program', http://www.fourmilab.ch/random/,  accessed Dec.  2017

\bibitem{19} Wen-Kai, Y.,   Shen, L.,  Xu-Ri Y., \textit{et al.}: `A protocol based on compressed sensing for high-speed
authentication and cryptographic key distribution over a
multiparty optical network', \textit{Appl. Optics}, 2013, \textbf{52}, pp. 7882--7888

\bibitem{20} `ID Quantque Random~Numbers',~http://www.idquantique.com/random-number-generation/quantis\,random\,number\,generator,  accessed Dec. 2017

\bibitem{21} `Embedded Device Hacking`,  http://www.devttys0.com/2013/06/differentiate-encryption-from-compression-using-math,  accessed Dec. 2017

\bibitem{25}  Handkiewicz, A.,  Katarzynski, P.,  Szczesny, S.,  \textit{et al.}: `Design automation of a lossless multiport network and its application to image filtering', \textit{Expert Syst. Appl.}, 2014, \textbf{41},   
  pp. 2211--2221

\bibitem{25a} Handkiewicz, A., Szczesny, S., Naumowicz, M.,  \textit{et al.}:  `SI-Studio, a layout generator of current mode circuits',  \textit{Expert Syst. Appl.}, 2015, \textbf{42}, pp. 3205--3018 

\bibitem{25b} Marszalek, W., Amdeberhan, T., Riaza, R.:  `Singularity crossing phenomena in DAEs: a two-phase fluid flow application case study', \textit{Comp. Math.  Appl.}, 2005, \textbf{49}, pp. 303-319

\bibitem{26} Campbell, S. L., Marszalek, W.: `Mixed symbolic-numerical computations with general DAEs II: An applications case study', \textit{Numerical Algorithms}, 1998, \textbf{19}, pp. 85-94

\bibitem{26a} Melosik, M.:  `Reconfigurable mixed-mode hybrid bit generator for chaos based cryptography', \textit{Ph.D. Thesis}, Poznan Univ. of Technology, Poland, 2017

\bibitem{27} Tehranipoor, M., C.  Wang, C. (Eds.), \textit{Introduction to hardware security
and trust} (Springer, New York. 2012)

\end{thebibliography}
\end{document}